# MVDC Power Cables for Future Wide-Body All-Electric Aircraft: A Research Journey from System-Level Requirements to Modeling, Design, Fabrication, and Testing

Babak Porkar, *Senior Member, IEEE* and Mona Ghassemi, *Senior Member, IEEE*
Department of Electrical and Computer Engineering
The University of Texas at Dallas
Richardson, TX, USA

***Abstract*—**Future wide-body all-electric aircraft require lightweight electrical distribution systems capable of delivering megawatt-scale propulsion power under demanding thermal and electrical conditions. This article reviews our research program on medium-voltage direct-current (MVDC) power cables, tracing its progression from aircraft-level requirements to coupled modeling, cable design, prototype fabrication, and low-pressure electrical testing. The review focuses on publications arising from this program to explain the connections between successive investigations and the decisions guiding their development. Aircraft-level power-flow and single-contingency studies establish a ±5 kV distribution architecture and a 1 kA design objective for the most heavily loaded cable circuits. Coupled electrical, thermal, and airflow models subsequently examine multilayer multifunctional electrical insulation, conductor geometry, bipolar pole spacing, duct configuration, and forced convection. Flight-phase investigations distinguish the higher-current takeoff condition from lower-current operation at reduced cruise pressure. Experimental studies progress from bonded flat specimens to wrapped cable prototypes, examining fabrication conditions, partial discharge, and dielectric breakdown at atmospheric pressure and pressures down to 18.8 kPa. The findings demonstrate the importance of insulation interfaces, wrapping method, cable geometry, and installation boundaries in achieving compact designs with improved electrical performance. Together, these investigations provide an integrated development framework connecting aircraft requirements with manufacturable cable prototypes. The distinction between predicted current-carrying capability and measured insulation performance is maintained throughout, identifying the combined electrical, thermal, mechanical, and endurance evaluations still required for aircraft qualification.

## I. Introduction

Battery-electric propulsion and hydrogen-based propulsion are two pathways to reducing aviation's reliance on conventional jet fuel. A comparison of these pathways in terms of energy storage, efficiency, infrastructure requirements, and environmental impact is provided in [1]. Battery-electric propulsion is best suited to shorter missions because the low specific energy of current batteries limits the range achievable within aircraft mass constraints. Hydrogen offers a potential pathway for longer missions, although storing liquid hydrogen onboard and supplying it at airports pose substantial challenges. Hydrogen can power aircraft through combustion or through fuel cells that generate electricity for propulsion motors. The fuel-cell route converges with battery-electric propulsion through their shared use of electric motors, as illustrated in

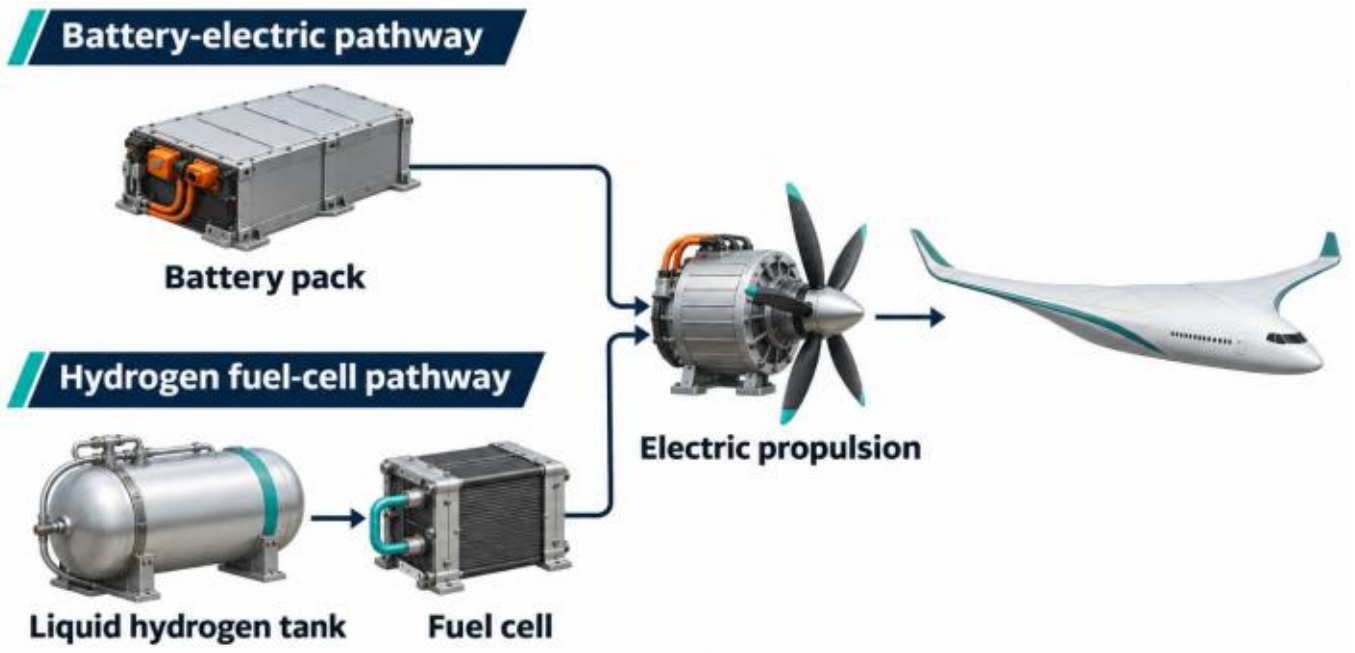

Fig. 1. Battery-electric and hydrogen fuel-cell pathways for aircraft propulsion. Batteries supply electricity from onboard storage, while fuel cells convert stored hydrogen into electricity. In both pathways, electric motors provide propulsion.

Fig. 1. Both battery-electric and hydrogen fuel-cell aircraft therefore require onboard electrical distribution systems sized to deliver propulsion power.

Narrow-body commercial aircraft have a single passenger aisle, whereas wide-body aircraft typically have two. As shown in Fig. 2, the Airbus A320neo and Boeing 737-8 illustrate the narrow-body class, with published seating ranges of 150–180 and 160–180 and published flight ranges of 3,400 and 3,500 nmi, respectively [2, 3]. The wide-body Boeing 787-9 and Airbus A350-900 seat 250–325 and 300–350 and have published ranges of 8,300 and 8,600 nmi, respectively [4, 5]. nmi stands for nautical mile, a unit of distance used in aviation and marine navigation. 1 nmi is exactly 1.852 km, or about 1.15 miles. The two narrow-body examples have published altitude ceilings of 39,800–41,000 ft, while both wide-body examples have a published ceiling of 43,100 ft [6]. These examples illustrate the scale of the wide-body mission in terms of passenger capacity and range, as well as the high-altitude environment relevant to onboard electrical insulation.

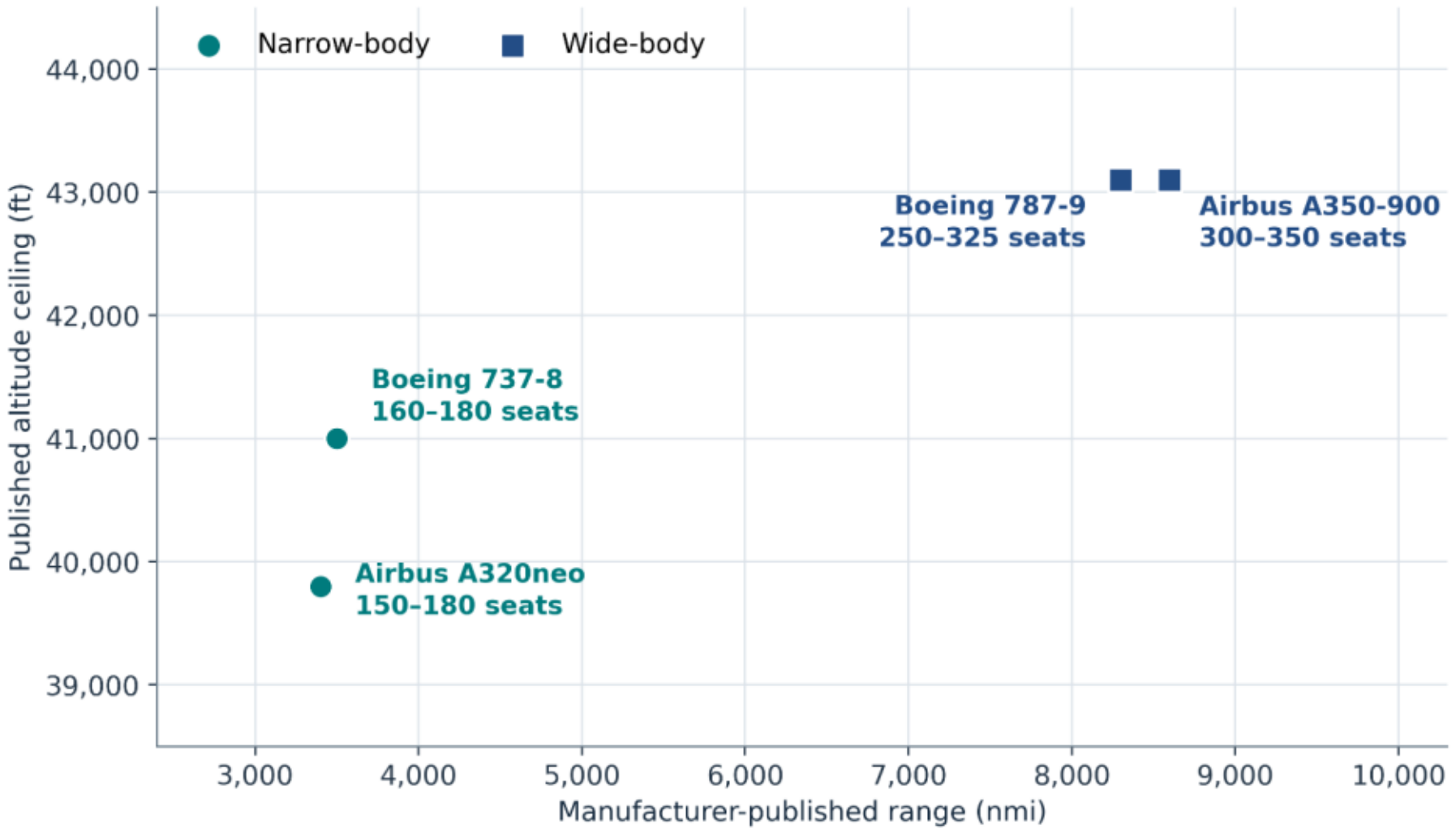

Fig. 2. Manufacturer-published range, published altitude ceiling, and seating ranges for representative narrow-body and wide-body aircraft [2–6]. Range and ceiling describe separate performance limits.

Aircraft electrification changes both the function and the scale of the onboard electrical power system. In more-electric aircraft (MEA), electrical systems take over functions traditionally served by hydraulic or pneumatic systems, such as actuation, environmental control, and ice protection, while propulsion remains fuel-driven. In all-electric aircraft (AEA), the electrical power system also supplies propulsion, making it responsible for delivering tens of megawatts during takeoff in large passenger-aircraft concepts. Our earlier review of MEA and AEA power systems examined how this transition affects power distribution architectures, voltage levels, and power flow [7]. In a subsequent review focused on power-system components, we considered a conceptual all-electric adaptation of the NASA N3-X with an approximately 25 MW takeoff power requirement and a bipolar ±5 kV DC propulsion bus (10 kV pole-to-pole) [8].

By late 2026, electric propulsion had reached megawatt-class flight demonstrations, while fully electric propulsion for wide-body aircraft remained at the conceptual design stage. In Europe, Airbus and MTU Aero Engines, a Munich-based German aircraft-engine manufacturer whose initials derive from *Motoren- und Turbinen-Union*, announced plans for a joint venture to develop and commercialize a propulsion system in which hydrogen fuel cells supply electricity to drive the propellers [9]. In the United States, GE Aerospace, NASA, BETA Technologies, and Boeing demonstrated megawatt-class hybrid-electric propulsion above 30,000 ft on a modified Saab 340B [10]. At the regional scale, U.S.-based BETA Technologies' battery-electric ALIA CX300 completed 23 trial flights with Scottish airline Loganair on routes linking six Scottish cities, illustrating progress toward battery-electric short-haul operations [11]. At the wide-body scale, a recent design study evaluated hydrogen fuel-cell electric propulsion for missions representative of the Boeing 787 and 777 and estimated motor power requirements of approximately 8–9 MW per engine [12]. A central design challenge for these aircraft is to deliver propulsion power within aircraft mass limits while maintaining the insulation integrity of cables and terminations at cruise altitude.

For aircraft power cables, the flight profile determines both the outside-air pressure and the duration of exposure. In the U.S. Standard Atmosphere, static pressure is approximately 23.8 kPa at 35,000 ft and 18.8 kPa at 40,000 ft [13]. Fig. 3 considers the highest published range in each aircraft class shown in Fig. 2. Dividing 3,500 nmi by an assumed mean cruise ground speed of 450 kt gives 7.8 h for the narrow-body Boeing 737-8. Dividing 8,600 nmi by 490 kt gives 17.6 h for the wide-body Airbus A350-900. The respective published altitude ceilings of 41,000 and 43,100 ft correspond to standard-atmosphere pressures of 17.9 and 16.2 kPa. Each screening case assigns the ceiling pressure to the entire range-derived equivalent cruise time. These paired assumptions define idealized exposure cases. In service, distance and ground speed vary by mission, and pressure changes during climb, cruise, and descent. Cables and terminations in vented, unpressurized bays can experience pressures close to outside-air pressure, whereas the local pressure in other compartments depends on the installation. The pressure and exposure duration are relevant to evaluating discharge behavior in air-filled regions and conductor heating under reduced convective cooling.

A published simulation of an Airbus A350-900 flight shows approximately 25 min for initial climb and 35 min for final descent, or about 1 h combined [14, Fig. 7]. Fig. 4 represents this realistic combined duration using 30 min for each phase. Fig. 4 uses 17.6 h as the total airborne duration of an illustrative mission. Cruise, including the step climb, occupies 16.6 h, or 94.3% of airborne time. The step climb from 35,000 to 40,000 ft reduces standard-atmosphere outside-air pressure from approximately 23.8 to 18.8 kPa. The aircraft spends 10.5 h at 40,000 ft in this example. For cables in vented, unpressurized bays, the time spent at each pressure level is more informative than total cruise time alone.

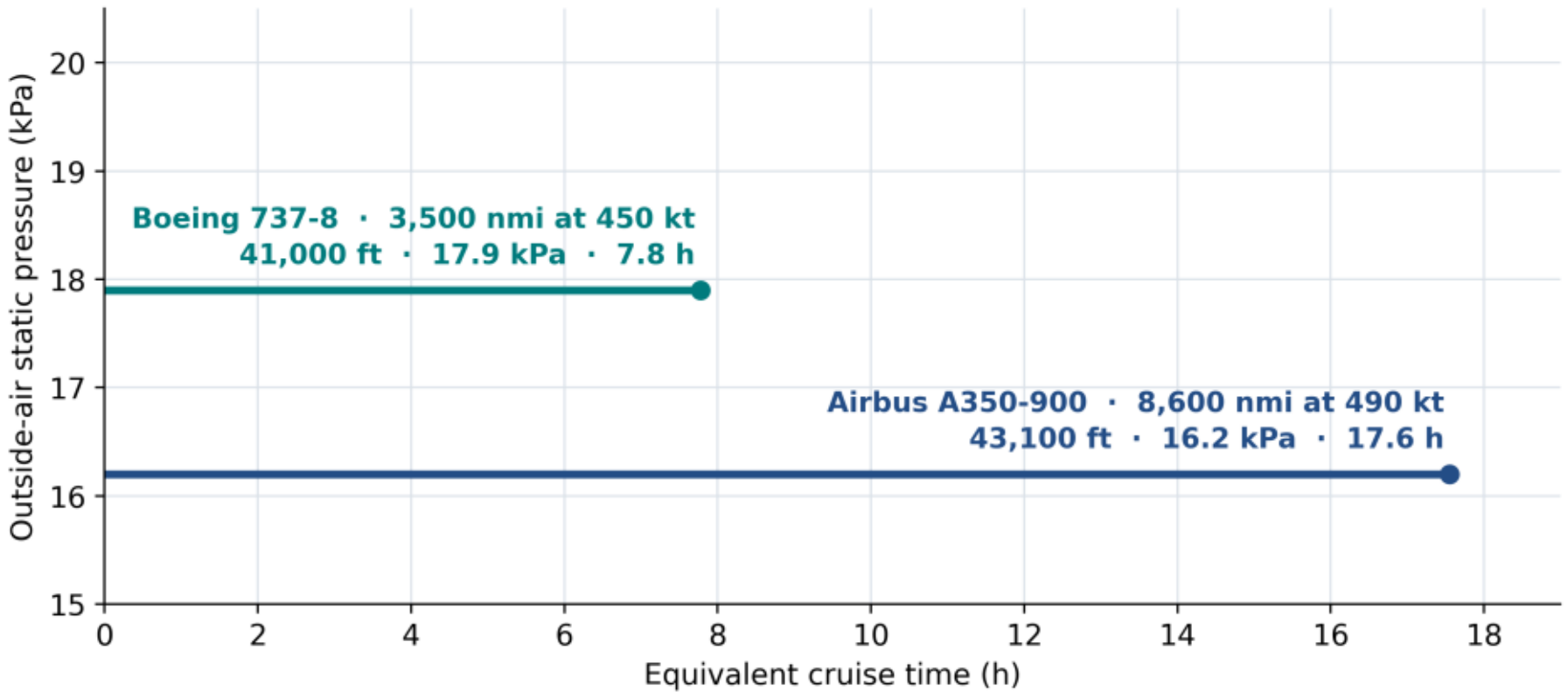

Fig. 3. Idealized low-pressure exposure for the maximum published ranges shown in Fig. 2. Equivalent cruise time is published range divided by assumed mean cruise ground speed. Each line holds the standard-atmosphere pressure at the published altitude ceiling for the full estimated duration. The cases combine separate published limits for screening purposes and do not represent measured flight profiles or verified operational bounds.

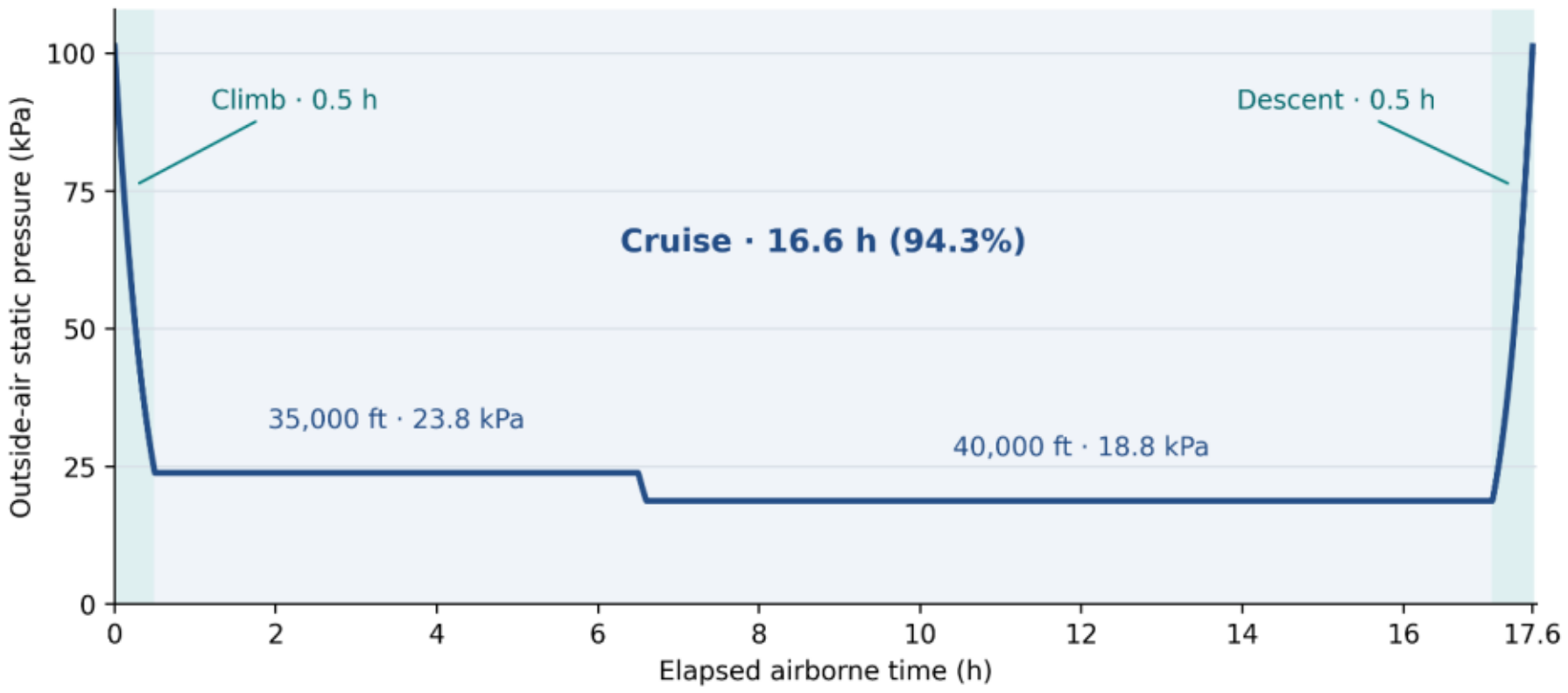

Fig. 4. Standard-atmosphere outside-air static pressure during an illustrative 17.6 h airborne mission. The combined climb and descent duration is informed by a simulated A350-900 profile [14] and is represented here as 30 min per phase.

The pressure surrounding an installed cable depends on its route through the aircraft. The passenger cabin and cargo compartments within the pressure vessel are pressurized, while equipment spaces outside this boundary can experience pressures close to outside ambient. The Airbus A350 and Boeing 787 maintain cabin pressures corresponding to an altitude of approximately 6,000 ft [5], [23]. This cabin altitude corresponds to a standard-atmosphere pressure of approximately 81 kPa [13]. By comparison, a cable in a vented, unpressurized compartment at 40,000 ft may experience approximately 18.8 kPa. Electrical equipment bays must be assessed individually because their pressurization depends on the aircraft layout. Consequently, the outside-air profiles in Figs. 3 and 4 apply to cable sections whose local pressure follows ambient conditions. Cable sections within pressurized compartments require the corresponding compartment pressure profile.

For a given transmitted power, increasing the distribution voltage reduces the current and can reduce conductor mass and resistive losses. However, higher voltage places greater demands on cable insulation and terminations, particularly under the reduced air pressure encountered at cruise altitude. This paper traces our work on the design, fabrication, and testing of medium-voltage direct-current (MVDC) power cables for wide-body all-electric aircraft. We begin with the aircraft-level propulsion power requirement and distribution architecture, then show how the choice of bus voltage and the allocation of power among feeders establish the voltage and current requirements for individual cables. We then describe how these requirements guide cable design, followed by the fabrication and experimental testing of cable prototypes.

## II. Propulsion Power Requirements and MVDC Distribution Architecture

The system studies behind the cable development began with steady-state analysis of MVDC networks. At the 2021 IEEE Workshop on the Electronic Grid, we presented Z-bus and monotone-mapping power-flow solvers and evaluated them on the proposed all-electric N3-X network [15]. At the 2022 IEEE Aerospace Conference, one paper applied monotone mapping to a bipolar ±5 kV DC propulsion network under normal operation and single contingencies [16], and a companion paper compared alternative propulsion distribution architectures [17]. These studies culminated in the journal paper [18], which compared three architectures under both ±5 kV DC and 10 kV AC operation and provides the system-level basis for the cable requirements discussed here. A broader discussion of electrical system architectures is provided in [19].

*A. Takeoff power and bus voltage*

NASA's N3-X is a long-range hybrid-wing-body concept with turboelectric distributed propulsion (Fig. 5). In this concept, propulsion fans are distributed across the upper rear of the aircraft. They ingest slower-moving air from the boundary layer over the airframe, an arrangement intended to improve propulsive efficiency. NASA's baseline design study used 14 propulsors and sized the aircraft to carry 300 passengers over 7,500 nmi at Mach 0.84 [20]. NASA's current overview describes a 16-fan configuration [21]. The 14-propulsor baseline is the configuration used in our all-electric power-system study [18]. Fig. 6 traces the power path in NASA's turboelectric baseline. Each wingtip turboshaft is a gas turbine that converts fuel energy into shaft power. Its output shaft turns superconducting electrical generators in the wingtip nacelle. The generators convert shaft power into electrical power, which the onboard distribution network carries to superconducting motors near the rear of the aircraft. Each motor converts electrical power back into shaft power and turns its local fan, which accelerates air to produce thrust. The electrical network thus links the wingtip generators to the rear motors and allows their locations and speeds to be selected separately [20], [21].

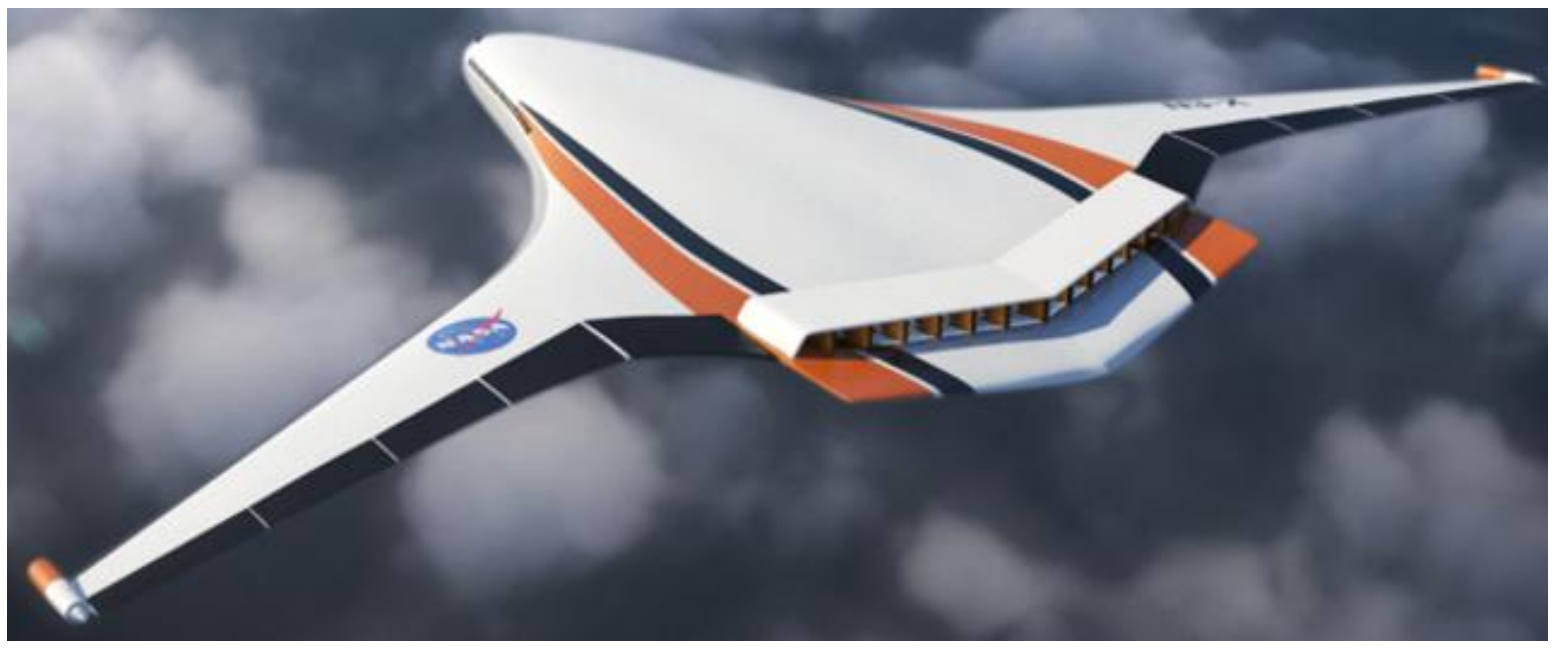

Fig. 5. NASA artist's rendering of the turboelectric N3-X concept, showing the hybrid-wing-body airframe and aft distributed fans [21].

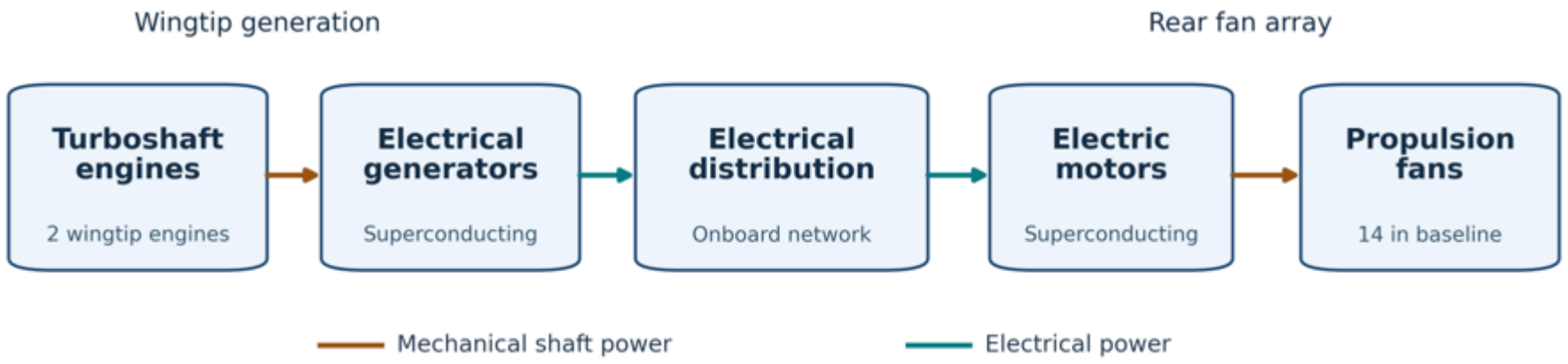

Fig. 6. Functional power path of NASA's 14-propulsor turboelectric N3-X baseline [20]. Each block represents a function, and the arrows indicate the form of power transferred.

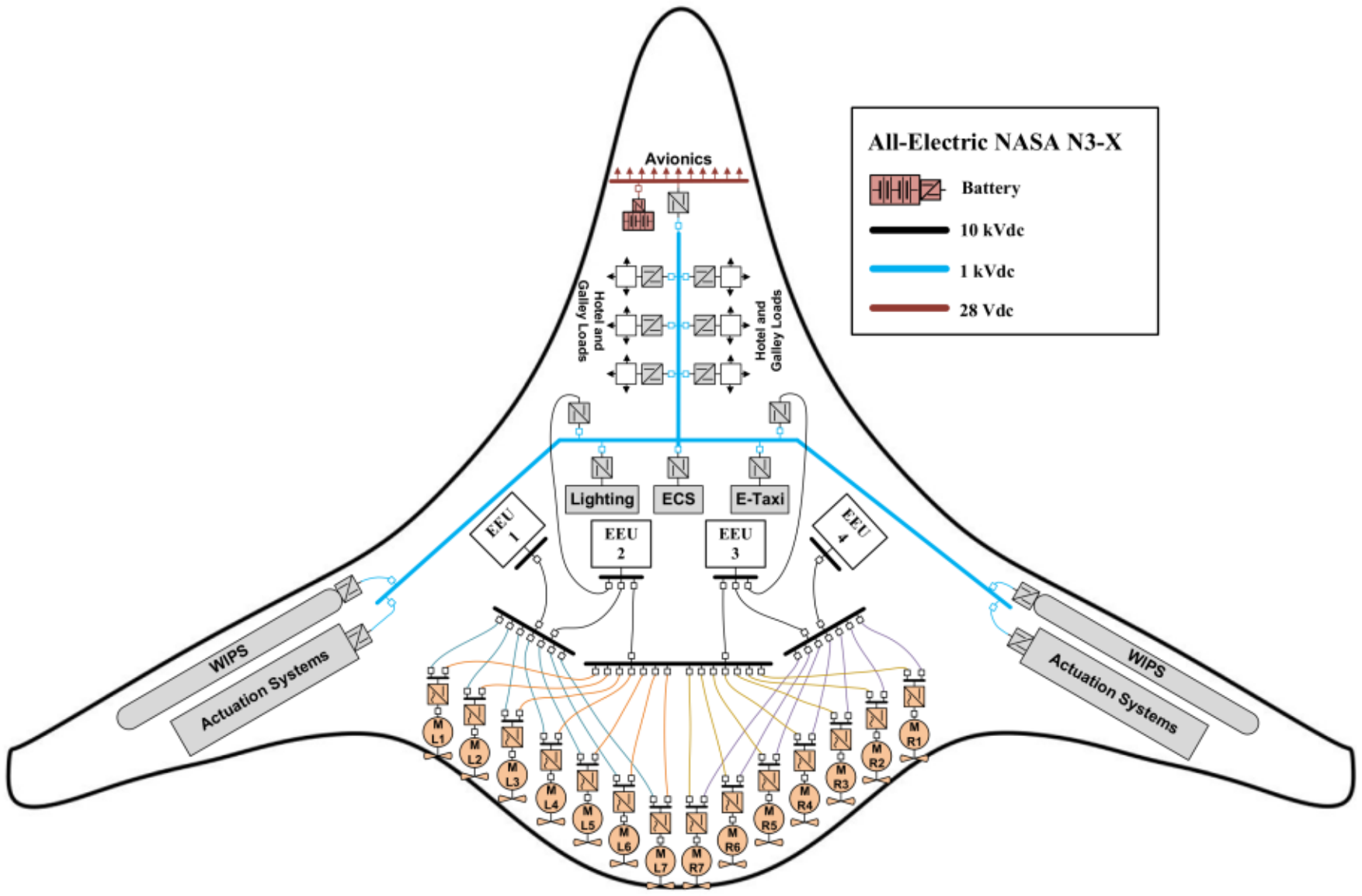

Fig. 7. Proposed electrical power system for the all-electric NASA N3-X concept [18].

As of November 2026, NASA continued to list the N3-X as a conceptual aircraft design [21]. Public NASA materials reviewed for this paper report neither an N3-X prototype flight nor a flight-test schedule. Related electrified-propulsion research has advanced through separate efforts, including NASA's AACES 2050 conceptual design studies presented in 2026 [22] and the 2026 megawatt-class hybrid-electric flight demonstration on a modified Saab 340B [10]. In the all-electric adaptation shown in Fig. 7, four onboard electrochemical energy units (EEUs) supply electrical power in place of the turboshaft–generator sets. The distribution network still delivers electrical power to 14 motor–fan sets. This change replaces the mechanical-to-electrical generation stage at the source while retaining the electrical-to-mechanical conversion at each propulsor. The takeoff power-flow model represents each propulsion motor as a 1.785 MW constant-power electrical load at its supply bus. The aggregate modeled propulsion demand is therefore 24.99 MW, or approximately 25 MW. This is the electrical demand assigned to the propulsion loads in the model. The sources also supply the modeled distribution losses, and nonpropulsive loads are excluded from the takeoff power-flow cases [18]. Fig. 7 places the three DC voltage levels in the context of the aircraft layout. The four EEUs feed the 10 kV pole-to-pole propulsion network, which distributes power to 14 motor–fan sets along the aft edge of the airframe. The 1 kV pole-to-pole network supplies nonpropulsive loads, including wing ice protection, actuation, and environmental control, while the 28 V DC network supplies avionics. The takeoff power-flow analysis focuses on the propulsion network [18]. We selected a bipolar ±5 kV DC bus as a design point within the ±4.5–12 kV DC range discussed in the NASA trade studies reported in [18]. The selected voltage is 10 kV between poles and 5 kV from either pole to the midpoint in magnitude under nominal balanced conditions. The pole-to-pole voltage governs the current for a load connected across the two poles, whereas the conductor-to-shield or conductor-to-airframe stress depends on how the shield and airframe are referenced. The study used 10 kV pole-to-pole as a practical point for comparing architectures. A broader voltage optimization would require component-level models for the cables, converters, and protection equipment [18].

*B. Propulsion distribution architectures*

Figs. 8–10 show the three 14-motor propulsion architectures studied in [18] at ±5 kV DC (10 kV pole-to-pole). Each uses four electrochemical energy units connected at source buses 1–4. The source blocks show battery and fuel-cell branches with power converters, together with supercapacitor branches. Downstream feeders connect the distribution busbars to motor buses, where inverters supply the motor–fan sets. ML and MR identify the left- and right-side motors. The distribution layouts determine which cables can continue supplying the motor buses following isolation of a source, busbar, or cable [18]. Architecture 1 (Fig. 8) has three distribution busbars, numbered 5–7. Source buses 1 and 4 connect to the outer busbars 5 and 7, respectively. Source bus 2 connects to busbars 5 and 6, and source bus 3 connects to busbars 6 and 7. The 14 motor buses are numbered 8–21. If EEU2 is disconnected while bus 2 remains available, power can flow from busbar 5 through bus 2 to busbar 6, transferring load to feeders that carry less current in normal operation [18]. Architecture 2 (Fig. 9) also uses three distribution busbars and motor buses 8–21. Its source-side connections put EEU1 on busbar 5, EEU2 and EEU3 on busbar 6, and EEU4 on busbar 7. The entire modeled 6.664 MW output of EEU2 in normal operation passes through the single feeder 2–6, corresponding to approximately 666 A at nominal voltage. Under loss of EEU4, the modeled current in motor feeder 6–15 rises to approximately 221 A, above the 215 A catalog ampacity of the Pansy conductor used in that case.

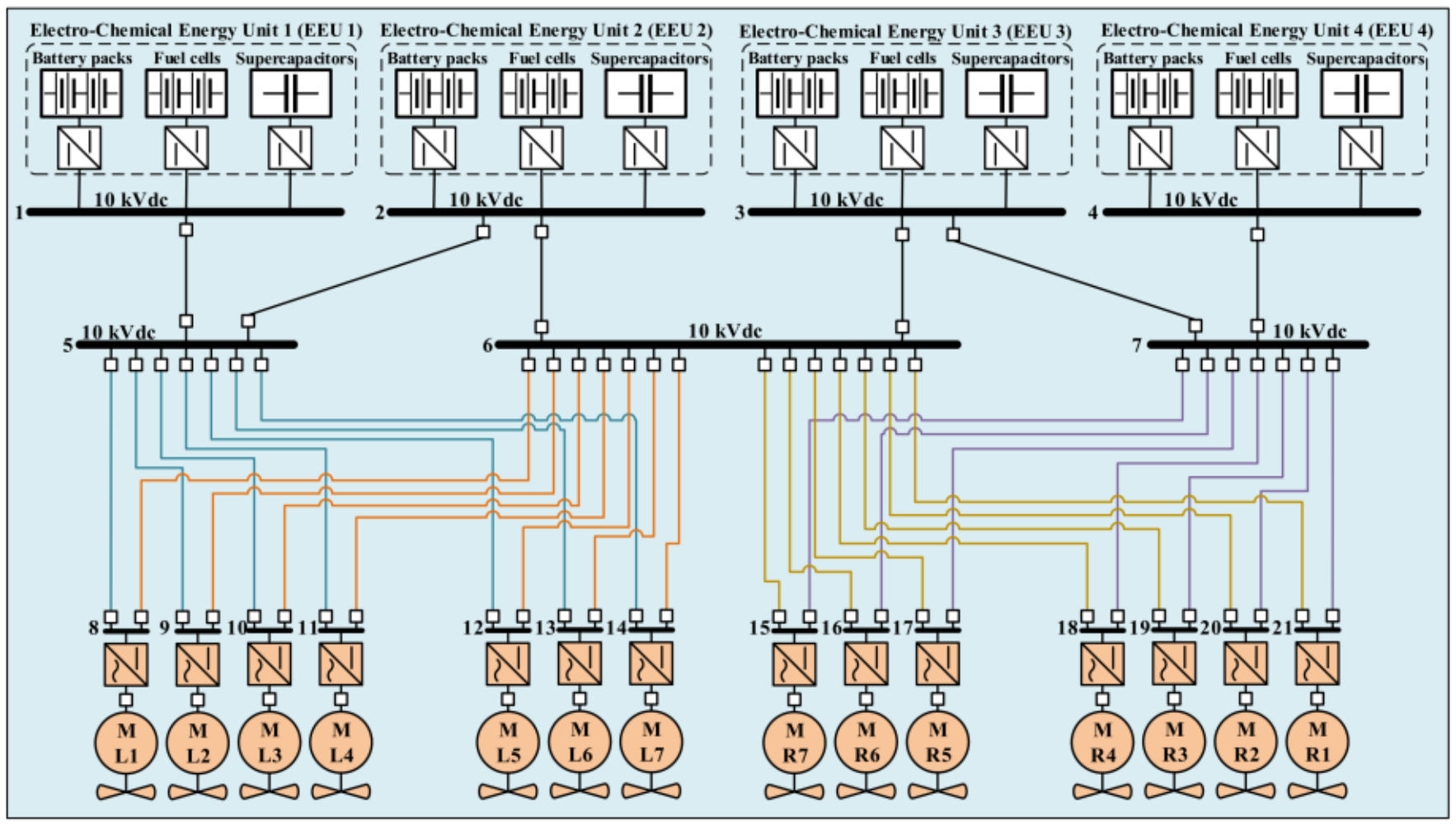

Fig. 8. Architecture 1 with three propulsion busbars and 14 motor–fan sets [18].

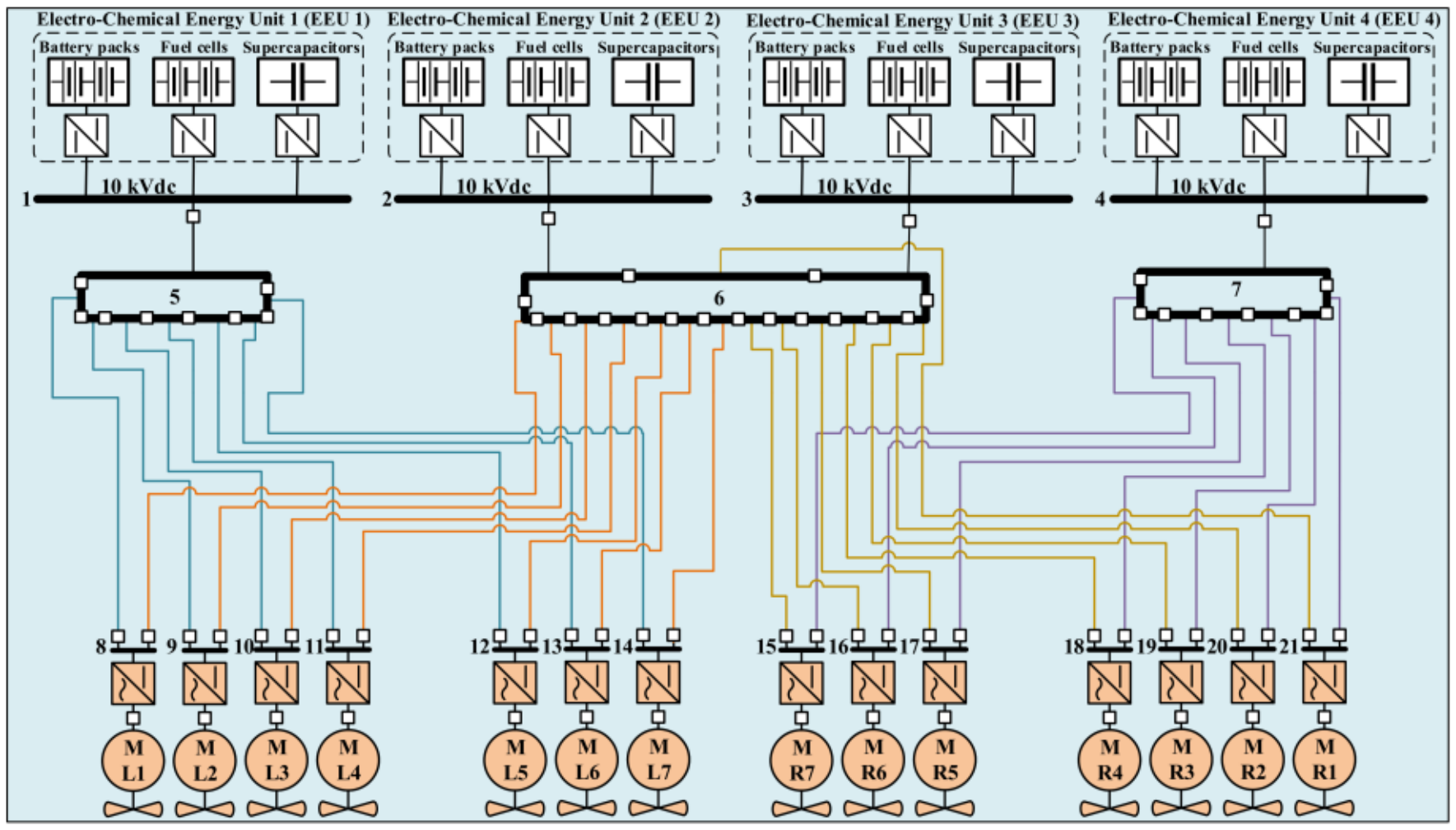

Fig. 9. Architecture 2 with three propulsion busbars and 14 motor–fan sets [18].

This layout has the shortest source-side cable length and the fewest counted breakers among the three alternatives [18]. Architecture 3 (Fig. 10) uses four distribution busbars, numbered 5–8, and motor buses 9–22. Two feeder connections from each source bus distribute source current across more branches. In the modeled normal operating case, the maximum source-side feeder current is approximately 354.5 A. This layout has the lowest modeled normal-state DC losses for both conductor pairs considered in [18]. Its additional busbar, source-side cabling, and breakers increase the installation burden relative to Architecture 1. Fig. 11 shows a 16-motor variant of Architecture 1. It retains the same arrangement of four source buses and three distribution busbars and adds one motor–fan set on each side. At the same assumed 25 MW total propulsion demand, equal allocation gives about 1.563 MW per motor and approximately 156 A per motor feeder at nominal voltage. The comparative power-flow results below refer to the three 14-motor cases [18]. Helens and Mazama are the source-side conductor candidates, while Pansy and Poppy are the motor-feeder conductor candidates evaluated in [18]. Table I compares the three 14-motor layouts. The source-side cable length increases from 68.46 m in Architecture 1 to 100.38 m in Architecture 3, while the total motor-side feeder length is 466.02 m for all three. The breaker counts follow the counting convention in [18], which excludes breakers at the energy-unit interfaces to source buses 1–4 and at the motor inverter interfaces common to all three cases.

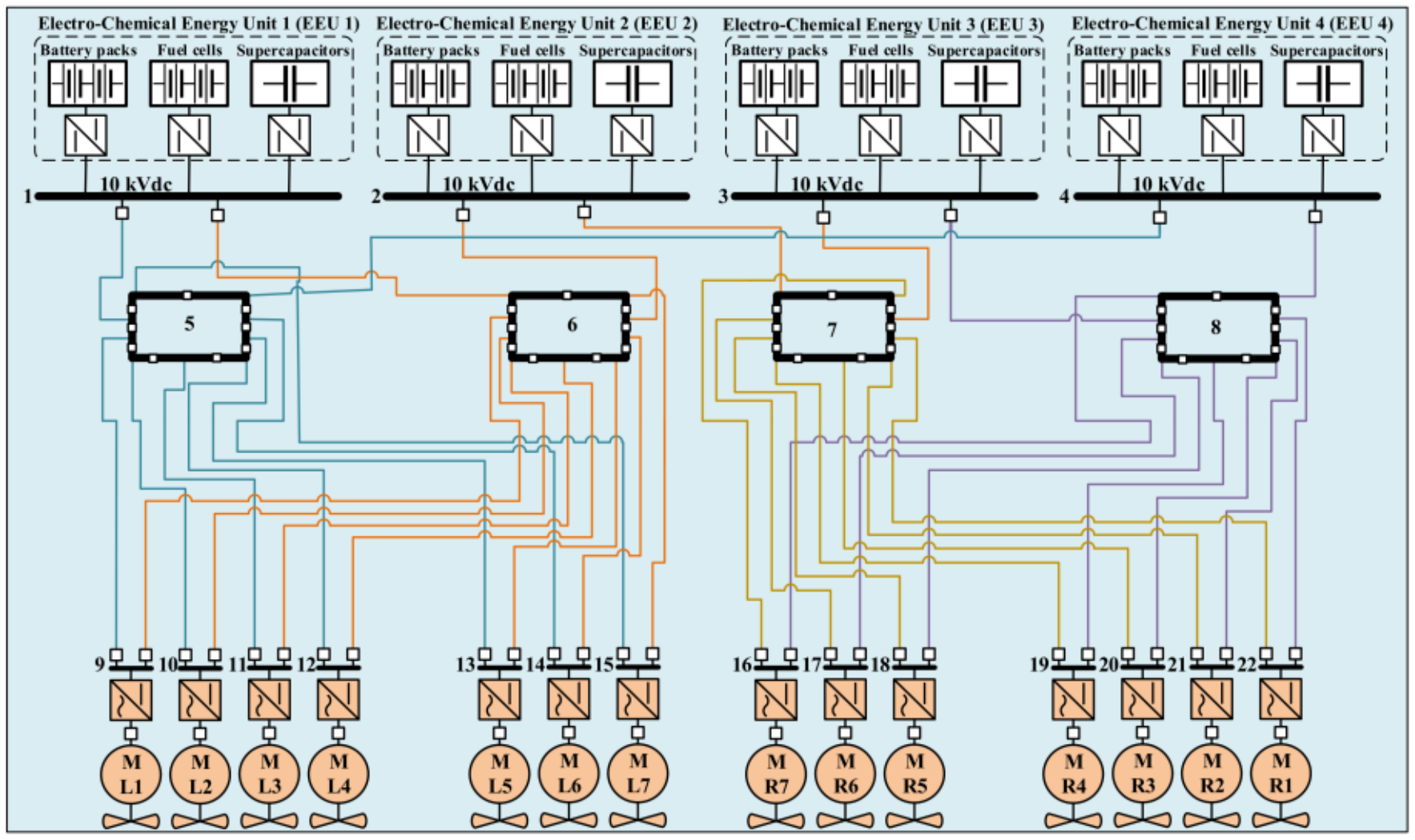

Fig. 10. Architecture 3 with four propulsion busbars and 14 motor–fan sets [18].

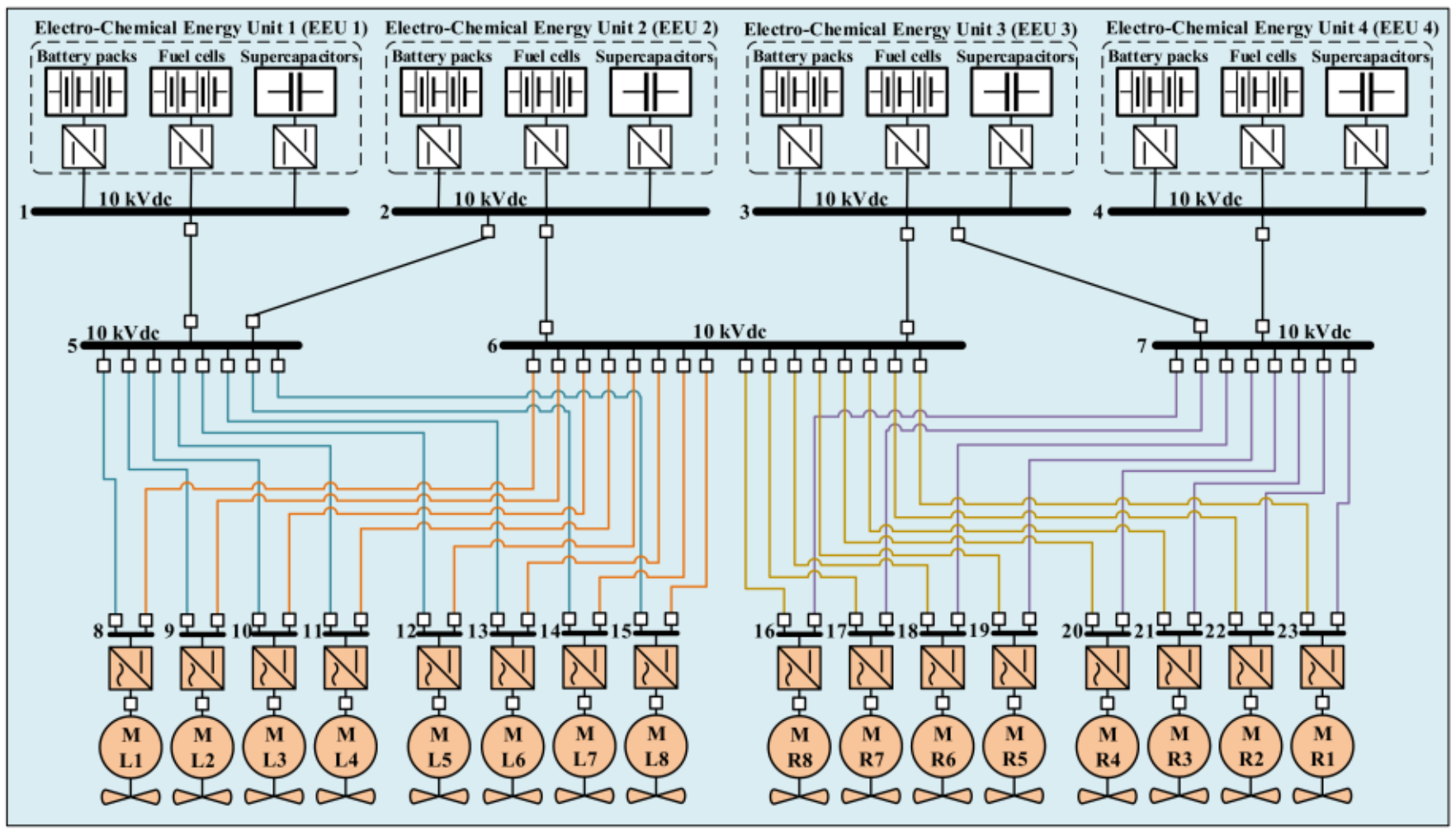

Fig. 11. Architecture 1 variant with 16 motor–fan sets [18].

TABLE I. COMPARISON OF 14-MOTOR PROPULSION ARCHITECTURES [18]

| Architecture | Distribution busbars | Source-side cable (m) | Circuit breakers | Normal DC loss (kW) Helens/Pansy \| Mazama/Poppy |
|---|---|---|---|---|
| 1 | 3 | 68.46 | 68 | 4.67 \| 3.75 |
| 2 | 3 | 50.19 | 64 | 4.81 \| 3.86 |
| 3 | 4 | 100.38 | 72 | 4.15 \| 3.30 |

The contingency study in [18] evaluates steady-state operation after the loss of one cable, busbar, or EEU. With the Mazama/Poppy pair, the highest branch loading across the reported DC contingencies is 79.23% in Architecture 1, 88.50% in Architecture 2, and 71.43% in Architecture 3, relative to the catalog ampacities. These percentages are sensitive to the conductor pair and to any derating applied for the installed environment. Reference [18] ranked Architecture 1 first because it combines acceptable modeled loading with fewer breakers, one fewer distribution busbar, and shorter source-side cabling than Architecture 3. The associated reliability and resilience considerations are discussed in [24].

*C. From network power flow to cable current requirements*

At nominal voltage, a dedicated feeder supplying one 1.785 MW motor has a current of 1.785 MW / 10 kV = 178.5 A in each pole conductor. The same calculation gives about 2.5 kA for the aggregate propulsion load at 10 kV. The actual current in any source feeder or interbus link depends on the network topology, source dispatch, and operating condition. For example, the loss of one energy unit leaves three to meet the demand, giving a first-order source allocation of about 8.3 MW per surviving unit before losses. This corresponds to about 830 A at 10 kV if the full output of one unit passes through a single feeder. The journal study used up to approximately 1 kA for preliminary screening of source-to-busbar conductors [18]. To determine branch currents, the journal study estimated cable lengths from the aircraft geometry, calculated DC resistances at a conductor temperature of 75 °C, and solved the network under normal operation and after the loss of a cable, busbar, or energy unit. For each branch, the required steady-state current is set by the largest current magnitude across the operating cases [18]. The resulting current requirements form the input to cable design. The catalog ampacities used to screen conductors were based on specified ambient temperature and airflow. The finished cable's ampacity depends on heat transfer through its insulation and the reduced convective cooling at flight altitude. The network model therefore establishes the current the cable must carry, while the cable design must establish whether it can carry that current at an acceptable conductor temperature and withstand the electrical stresses of the selected bus [18]. For the selected ±5 kV DC distribution system, the worst single-contingency operating cases, together with a modest design margin, led us to adopt a current requirement of 1 kA for the most heavily loaded cables [18]. This requirement applies to selected source-side feeders and interbus links, while individual motor feeders carry substantially lower currents. We therefore adopted ±5 kV DC and 1 kA as the target voltage and current ratings for the cable design, prototype fabrication, and experimental studies presented in the following sections. At the system-planning stage, the insulation structure and thermal performance of the aircraft cables had not yet been established. Bare-conductor data therefore provided preliminary conductor dimensions and resistances for calculating bus voltages and branch currents. The takeoff cases determined the maximum current demand under the modeled contingencies, while the complete cable's allowable current remained a cable-design requirement. We subsequently adopted a 1 kA current-rating target to accommodate the calculated demand with additional margin. The following cable-design studies assess whether this target can be achieved after accounting for insulation, installation conditions, and heat transfer.

## III. Cable Design from Low Pressure Thermal Limits to Flight Phase Requirements

Our review of insulation materials and systems for more- and all-electric aircraft examined partial discharge, arc tracking, thermal aging, material selection, and the limitations of high-altitude testing and existing standards [25]. A subsequent review focused on low-pressure insulation challenges in electric machines, power cables, and printed circuit boards, including discharge detection and mitigation [26]. Together, these reviews show why higher operating voltages and reduced pressure require coordinated electrical, thermal, and material design across the aviation powertrain, providing the broader context for the cable studies discussed below. Cable-specific insulation and thermal-management challenges are also discussed in [27]. The United States section of a recent global review also discusses these aircraft cable studies within broader progress in polymeric insulation [28].

A more recent review examined medium-voltage power cables specifically for electrified aircraft and spacecraft, connecting insulation physics with reduced-pressure discharge behavior, thermal management, space charge, and radiation-related degradation [29]. It also assessed the limitations of existing cable qualification approaches under combined environmental and electrical stresses. This broader assessment complements the design journey reviewed here: lightweight insulation must be evaluated as part of a complete cable system under its intended service environment, rather than selected from material dielectric-strength data alone.

With the system-level voltage and current requirements established, the next question was whether a complete insulated cable could carry the required current within an acceptable mass and diameter. Our first cable studies followed two connected directions. The 2022 IPMHVC conference paper examined the influence of reduced air pressure on a commercial 5 kV cable [30]; the more extensive IEEE Access study developed this analysis by investigating ambient temperature, enclosure geometry, radiative exchange, and temperature-dependent DC field redistribution [31]. The 2022 CEIDP paper then introduced a lightweight, high-temperature multilayer insulation design [32]. The subsequent IEEE Transactions on Dielectrics and Electrical Insulation journal study expanded the design approach into several multifunctional cable configurations addressing thermal management, partial discharge (PD), and arc tracking [33]. The journal studies provide the main quantitative results reviewed in this section, while the conference papers document the progression of the work. We subsequently linked the cable model to flight-phase power-flow results, comparing the higher-current takeoff condition with the lower-current, lower-pressure cruise condition. Two conference studies established this progression [34], [35], and the ensuing journal study brought the system analysis and bipolar cable optimization together [36].

These studies are numerical design investigations. Their results establish candidate geometries and predicted operating limits under specified boundary conditions; fabrication quality, discharge endurance, mechanical durability, and installed performance require experimental evaluation. The ±5 kV system rating is represented here by a single pole conductor at +5 kV relative to its grounded screen or prescribed external reference. The corresponding negative-pole cable must withstand the opposite polarity. The 1 kA requirement applies to each conductor of the heavily loaded bipolar circuit, which transmits 10 MW at 10 kV pole-to-pole under nominal balanced conditions [18], [33].

### *A. Why the bare-conductor assessment must become a complete-cable assessment*

The network model in Section II determines the current demanded by each branch using an assumed conductor resistance. Cable design must additionally determine the conductor temperature and the electrical stress in its insulation. Joule heating raises the conductor temperature; heat then travels through the insulation and jacket before leaving the surface through convection and radiation. Adding insulation increases the thermal resistance between the conductor and the cable surface, while reducing the cable's overall diameter decreases the surface area available for convective and radiative heat rejection per unit length. A conductor selected from a catalog therefore cannot be assigned the same current rating after insulation and installation without a corresponding thermal assessment [30]–[33]. The low-pressure calculations address cable sections installed in vented, unpressurized locations, as distinguished from pressurized compartments in Section I. Reference [30] used 16.8 kPa, whereas the two journal studies used 18.8 kPa, approximately the standard-atmosphere pressure at 12.2 km (40,000 ft) [31], [33]. Furthermore, the 1 kA current target arose from takeoff contingency demand, while the low-pressure studies apply that current at a cruise-representative pressure as a demanding cable-design condition. Section III.G explains how the later studies replaced this combined screening condition with distinct takeoff and cruise loading cases [34]–[36].

### *B. Coupled electrical, thermal, and airflow model*

Fig. 12 shows the computational geometry used in [33]. A single cable lies inside a square aluminum-alloy enclosure, with a nominal 1 in (25.4 mm) separation from the lower wall. A cylindrical air domain approximately 0.35 in (8.9 mm) thick surrounds the cable for the natural-convection calculation. The enclosing metal surface is included in the surface-to-surface radiation model. Its temperature and the prescribed external air-boundary temperature are held at the selected ambient value. The low-pressure baseline does not impose a forced cooling flow. The model couples heat conduction in the cable, buoyancy-driven laminar airflow, radiative exchange with the enclosure, and DC conduction through the insulation. The governing equations for each physical process and the coupling between them are detailed in [31]. Pressure affects air density and buoyancy-driven cooling, while the resulting temperature distribution changes conductor resistance and insulation conductivity. The latter determines how DC voltage is distributed across the insulation. Radiation supplies a parallel heat-rejection path whose effectiveness depends on surface temperatures, emissivities, and enclosure geometry [31], [33].

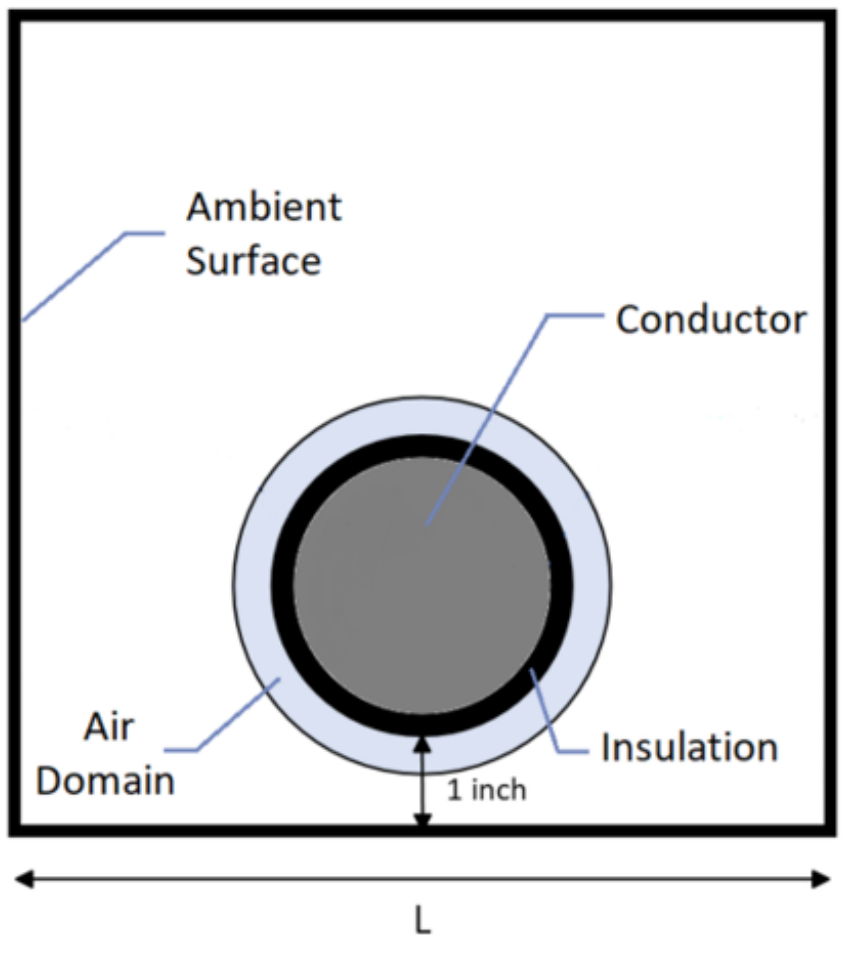

Fig. 12. Cable and enclosure geometry used for the coupled electrical–thermal–airflow analysis. L denotes enclosure side length [33].

### *C. Commercial-cable baseline: pressure derating and DC field inversion*

The initial studies examined a commercial 5 kV no-lead ethylene-propylene-rubber (NL-EPR) cable with a PVC jacket and a normal-operation temperature limit of 105 °C [30], [31]. Fig. 13 shows its predicted current capability at atmospheric pressure and 18.8 kPa. At 55 °C ambient, the permissible current decreased by approximately 14.8% for a 250 mm enclosure and 13% for a 1000 mm enclosure. Reduced natural convection caused this derating; the journal study reported a maximum convection reduction of approximately 58.3% at 25 °C. Radiation continued to remove heat, so the reduction in total current capability was smaller than the reduction in convective heat transfer [31].

Enclosure size and surface emissivity also influenced radiative cooling, demonstrating that ampacity depends on the installation as well as the cable construction [31]. The earlier conference study at 16.8 kPa predicted current reductions of approximately 15.5% and 13.8% at 40 °C for its two radiative-exchange configurations [30]. These values apply to their specified pressures, temperatures, and surrounding surfaces; they are not universal altitude derating factors.

Increasing current also raised the radial temperature gradient and the associated conductivity gradient across the insulation (Fig. 14). The hotter inner region became more conductive, shifting a greater share of the voltage toward the cooler outer region and intensifying DC field inversion. At 40 °C ambient, 18.8 kPa, and a 250 mm enclosure, the calculated maximum cable temperature rose from approximately 77.6 °C at 1000 A to 105 °C at 1300 A [31]. These findings established the need to assess thermal limits and insulation stress together.

### *D. First lightweight multilayer design*

The subsequent conference study [32] developed a cable for the ±5 kV, 1 kA target using an aluminum 1350 conductor and a thin, high-temperature multilayer insulation system. The stack combined an aluminum-nitride (AlN) inner dielectric layer with thermally enhanced Kapton MT+ polyimide (PI), perfluoroalkoxy (PFA) films, and a copper-tape screen. The design sought to offset the higher resistive losses of aluminum and the smaller heat-rejection surface of a compact cable by reducing insulation thermal resistance and allowing operation at a higher temperature.

At 1 kA, 40 °C ambient, and 18.8 kPa, the predicted overall diameter was 24.942 mm and the mass was 1.2169 kg/m. Relative to the modeled copper IEC-based cable, these values represented reductions of approximately 27.8% in diameter and 70.6% in mass; relative to the aluminum IEC-based cable, the reductions were 36.9% and 47.7%. The maximum calculated insulation field was 8.4 kV/mm. These comparisons used a 170 °C conductor-temperature limit for the multilayer design and 105 °C for the IEC-based designs, so the improvements reflect the combined material, geometry, and temperature choices [32]. This early AlN-containing concept led to the broader journal investigation of polymer-film configurations with additional functions [33].

### *E. Journal extension: insulation with multiple functions*

The journal study [33] extended the approach beyond thermal performance and dielectric strength. A cable for an aircraft must also manage surface discharge, arc tracking, and, where required, electromagnetic interference (EMI). Polyimide offers useful thermal and dielectric properties but can carbonize under damaging discharge conditions. Fluoropolymer-rich outer layers were therefore incorporated to improve resistance to arc tracking. Conductive screens were used to control the external electric field and provide EMI shielding; corona-resistant Kapton CRC layers provided an alternative means of improving discharge endurance without the mass of a metal screen. A corona barrier increases resistance to discharge damage; it does not remove the electric field from the surrounding air or guarantee PD-free operation [33].

The journal configurations used aluminum 1350 conductors and thin-film PI/PFA insulation. Kapton MT+ was selected for thermally optimized layers, Kapton HN provided a conventional PI comparison, and Kapton CRC provided the corona-resistant function. The thermal conductivities tabulated in [33] were 0.75 W/(m·K) for MT+, 0.20 for HN, 0.20 for CRC, and 0.195 for PFA. PFA and ETFE (ethylene-tetrafluoroethylene, sold as Tefzel) are distinct fluoropolymers; the PFA layers in the proposed designs should not be described as ETFE. Material-property fits and their temperature range remain part of the assumptions behind the predicted performance.

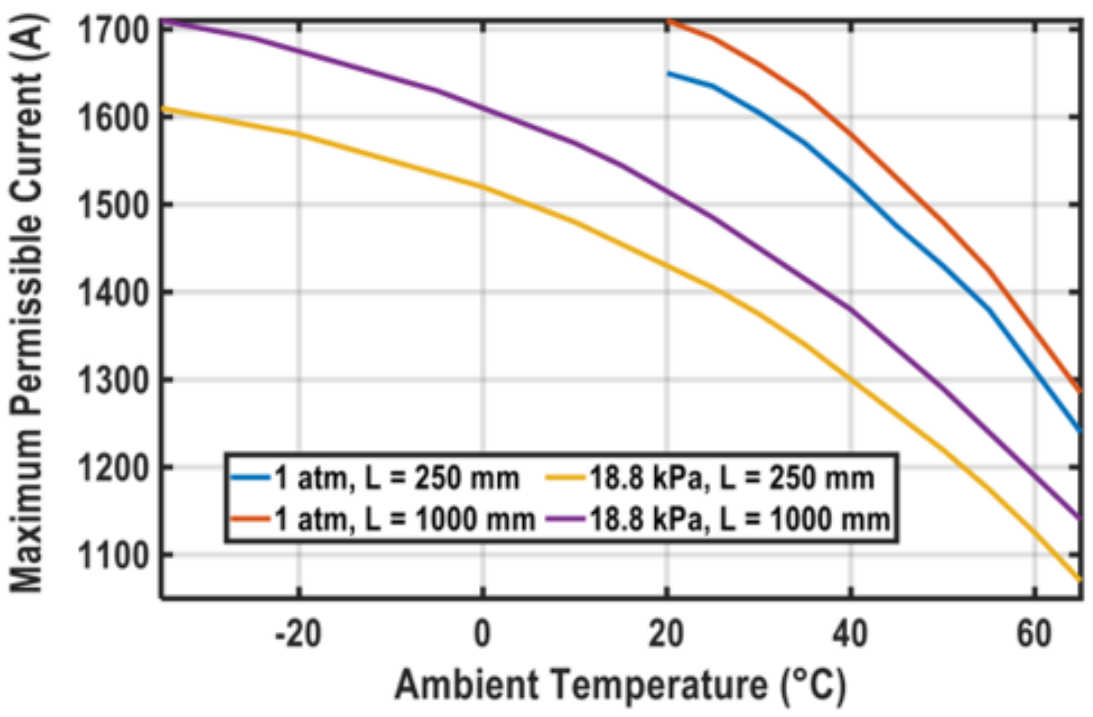

Fig. 13. Predicted maximum permissible current of the commercial NL-EPR cable as a function of ambient temperature, pressure, and enclosure side length, using a 105 °C thermal limit [31].

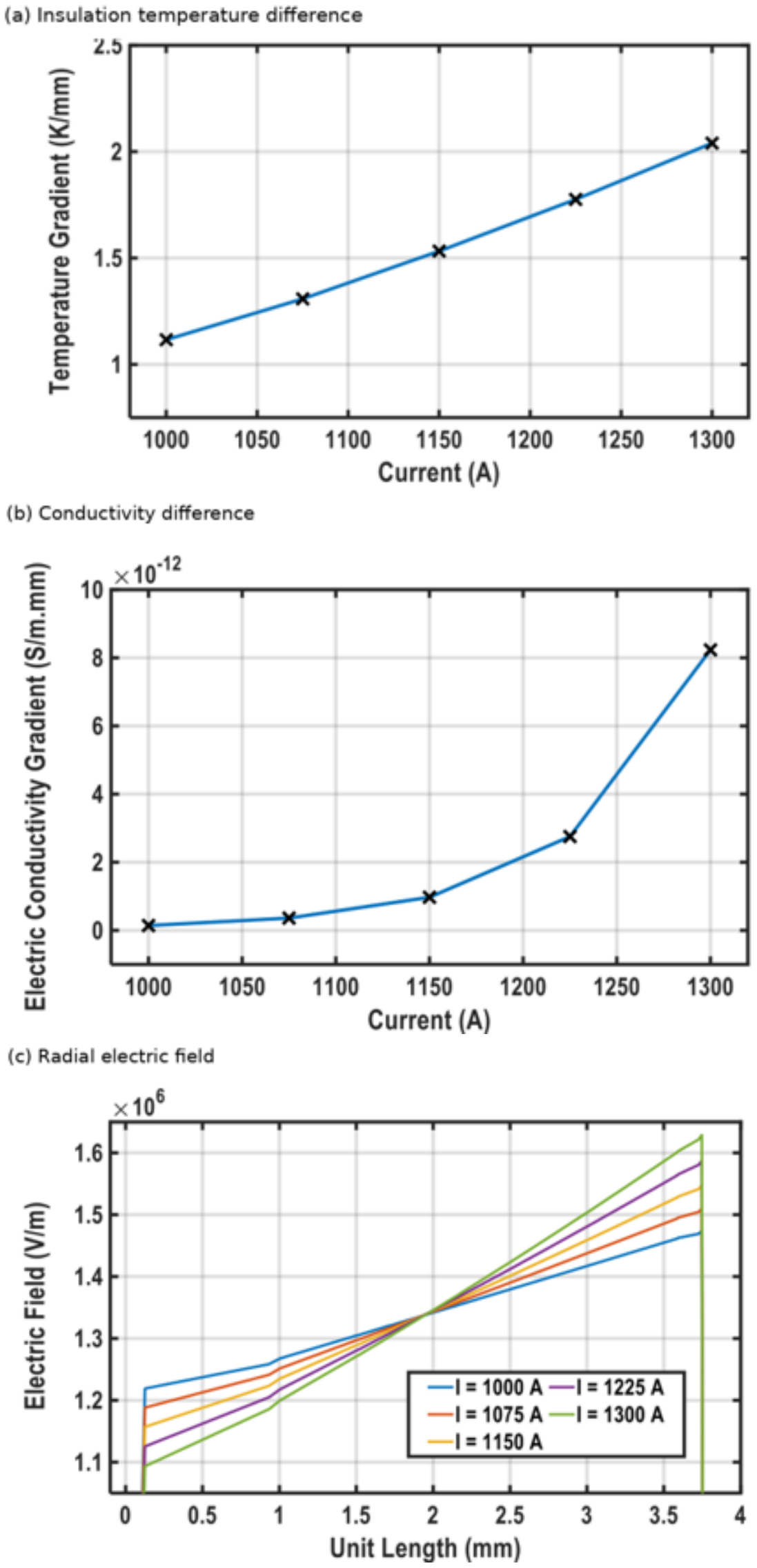

Fig. 14. Load-current dependence of the temperature and electrical-conductivity gradients across the insulation, and the radial electric-field distribution, at 40 °C, 18.8 kPa, and L = 250 mm, [31].

Figs. 15 and 16 retain all ten cross sections from the journal paper: six proposed configurations and four reference configurations. SC denotes a conductive screen, T denotes thermal optimization, PD denotes a corona-barrier function, and ARC denotes the added arc-tracking-resistant outer construction. MMEI denotes multilayer multifunctional electrical insulation. The configurations share common design principles but differ in which functions are provided and in the amount of copper or polymer added [33].

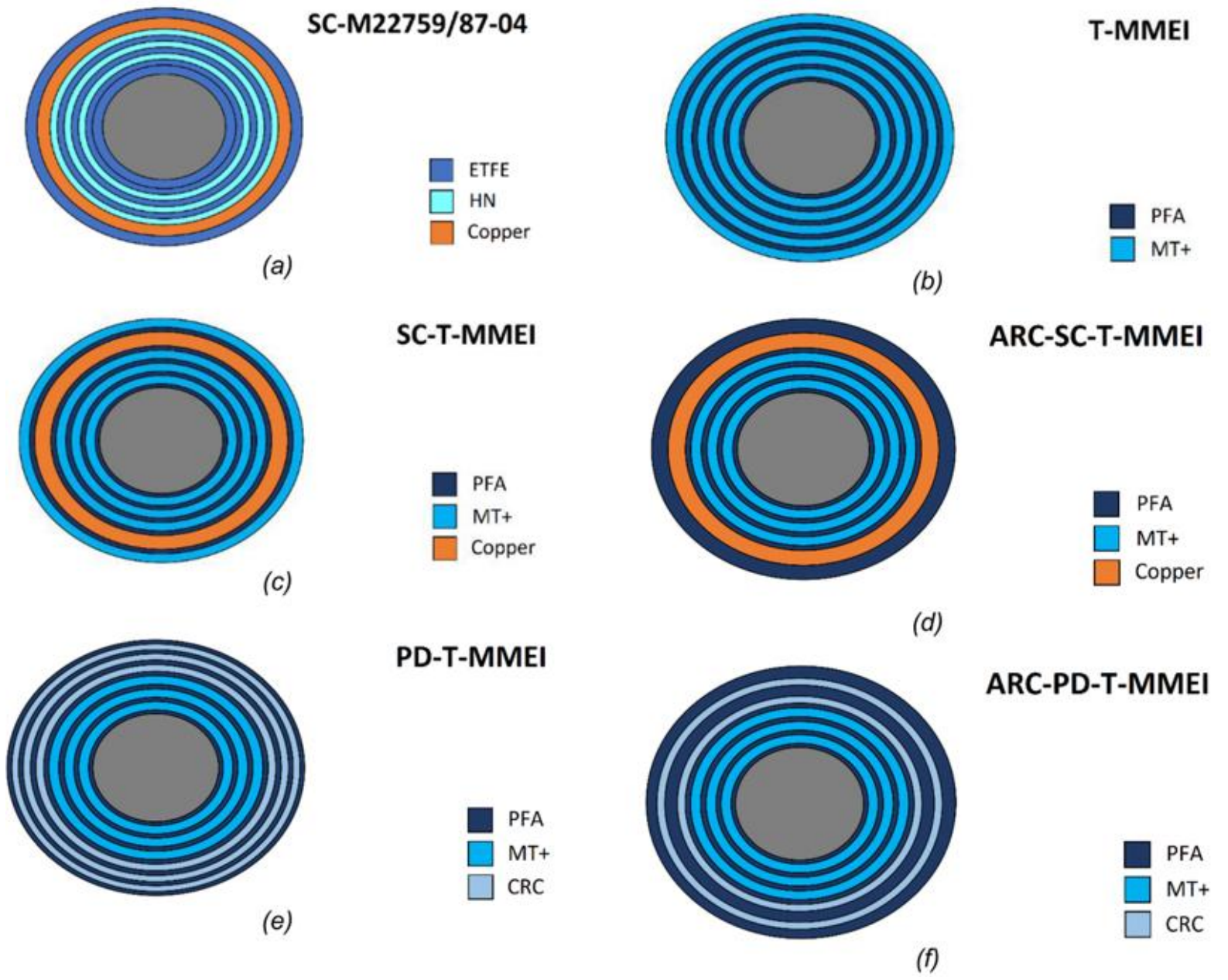

Fig. 15. Six proposed cable cross sections: (a) screened M22759/87-04 reference modification; (b) T-MMEI; (c) SC-T-MMEI; (d) ARC-SC-T-MMEI; (e) PD-T-MMEI; and (f) ARC-PD-T-MMEI [33].

T-MMEI uses a 12 mil (0.3048 mm) PI/PFA build composed of 1.5 mil MT+ and 0.5 mil PFA films. The other four thermally optimized variants begin with a 6.5 mil (0.1651 mm) MT+/PFA inner build. SC-T-MMEI adds a 5 mil (0.127 mm) copper screen and a 2 mil outer PI/PFA build. ARC-SC-T-MMEI uses the same screen with a 4 mil (0.1016 mm) PFA jacket. PD-T-MMEI adds a 4.5 mil (0.1143 mm) PFA/CRC outer build, while ARC-PD-T-MMEI uses a 6 mil (0.1524 mm) outer build with a larger PFA contribution. These are modeled radial builds and film gauges; actual wrapping overlap, bonding, void formation, and dimensional tolerances must be addressed during fabrication [33].

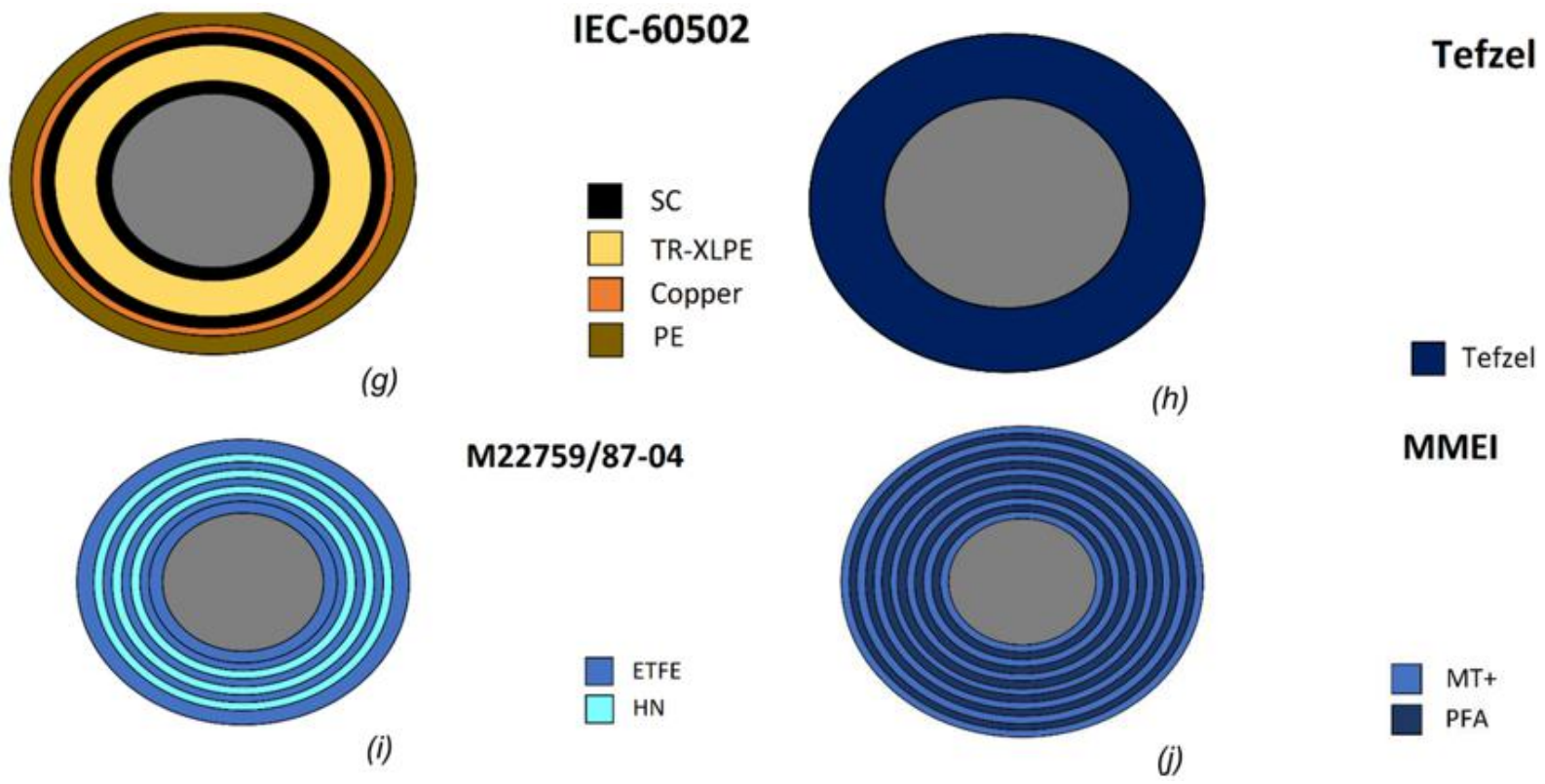

Fig. 16. Reference cross sections: (g) IEC 60502-based construction; (h) Tefzel; (i) M22759/87-04-based construction; and (j) MMEI. Reproduced from [33].

Table II summarizes the functions assigned in the source study. “Arc-tracking-resistant” describes the design intent and material choice, rather than a measured immunity to arcing. A screen controls the electric field around an intact cable body, but terminations, joints, screen interruptions, and defects remain separate discharge-critical locations. The AS22759-based reference constructions were adapted for numerical comparison at the study conditions; they should not be interpreted as certified MVDC aircraft cables [33].

TABLE II. INTENDED FUNCTIONS OF THE JOURNAL DESIGNS [33]

| Design | Thermally optimized | Surface-PD strategy | Arc-tracking-resistant outer design | EMI screen |
|---|---|---|---|---|
| T-MMEI | Yes | No added screen/barrier | No | No |
| SC-T-MMEI | Yes | Copper screen | No | Yes |
| ARC-SC-T-MMEI | Yes | Copper screen | Yes | Yes |
| PD-T-MMEI | Yes | CRC corona barrier | No | No |
| ARC-PD-T-MMEI | Yes | CRC corona barrier | Yes | No |
| SC-M22759/87-04 | No | Copper screen | Yes | Yes |

### *F. Design objective, constraints, and predicted dimensions*

The journal study compared designs using $J = m \times D$, the product of mass per unit length m (kg/m) and overall diameter $D$ (m). $J$ therefore has units of kilograms; Table III expresses it in grams. A smaller value favors a combination of low mass and small diameter, although $J$ is not itself a power-density measure. All compared designs were assessed at 1 kA, +5 kV conductor potential, 40 °C ambient, 18.8 kPa, and L = 250 mm. Conductor size was varied through repeated coupled simulations to identify the dimensions reported in the source [33].

The electrical constraint requires each dielectric layer's maximum field to remain below 80% of its adopted dielectric strength; the thermal constraints apply to every layer and the conductor. The design comparison used temperature limits of 105 °C for the IEC construction, 150 °C for Tefzel, and 260 °C for the other film-insulation constructions [33, Table I]. These are source-model design limits. Permitting a higher temperature reduces the conductor area needed to carry a given current under the assumed cooling conditions.

$$\text{minimize } J = m \times D$$
$$\text{subject to}$$
$$E_{max,i} < 0.8\, E_{bd,i}$$
$$T_{max,i} < T_{limit,i}$$
$$T_c < T_{limit,c}$$

TABLE III. PREDICTED 1 kA CABLE DIMENSIONS AND MASS AT 40 °C, 18.8 kPa [33, TABLE IV]

| Cable system | Core diameter (mm) | Cable diameter (mm) | Mass (kg/m) | J (g) |
|---|---|---|---|---|
| IEC 60502 | 27.41 | 39.64 | 2.3421 | 92.841 |
| Tefzel | 22.71 | 23.34 | 1.1342 | 26.472 |
| M22759/87-04 | 17.27 | 17.88 | 0.6669 | 11.924 |
| MMEI | 17.40 | 18.06 | 0.6755 | 12.200 |
| SC-M22759/87-04 | 17.25 | 18.11 | 0.7287 | 13.197 |
| T-MMEI | 17.22 | 17.83 | 0.6570 | 11.714 |
| SC-T-MMEI | 17.20 | 17.88 | 0.7107 | 12.707 |
| ARC-SC-T-MMEI | 17.15 | 17.93 | 0.7144 | 12.809 |
| PD-T-MMEI | 17.17 | 17.73 | 0.6522 | 11.564 |
| ARC-PD-T-MMEI | 17.20 | 17.83 | 0.6595 | 11.759 |

PD-T-MMEI had the smallest $J$ in the journal comparison: 11.564 g, with a 17.73 mm diameter and a mass of 0.6522 kg/m. Its $J$ was approximately 87.5% below the IEC-based comparator and 3.0% below the M22759/87-04-based comparator. ARC-PD-T-MMEI added the arc-tracking-resistant outer construction with a modest increase to 11.759 g, approximately 1.7% above PD-T-MMEI. PD-T-MMEI is therefore the minimum-$J$ option in this set, while ARC-PD-T-MMEI provides a broader combination of intended functions [33]. For screened constructions, SC-T-MMEI had $J$ = 12.707 g, about 3.7% below the screened M22759/87-04 reference. Adding the PFA jacket in ARC-SC-T-MMEI increased $J$ by approximately 0.8%, to 12.809 g. ARC-SC-T-MMEI combined the intended thermal, PD-control, EMI-shielding, and arc-tracking-resistance functions at a predicted diameter of 17.93 mm and a mass of 0.7144 kg/m [33]. Comparing like functions matters: a corona-barrier construction and a screened construction are not interchangeable where the installation requires shielding or external-field control.

The conference and journal results should also remain distinct. The AlN-containing conference candidate weighed 1.2169 kg/m under a 170 °C conductor limit [32], whereas the later journal candidates used different stacks and temperature assumptions. Their substantially smaller masses are not a direct same-temperature comparison of the AlN and polymer-only insulation systems. The journal results demonstrate the benefit of the full design choices within the numerical model, rather than establishing a general mass reduction for every multilayer cable.

The coupled electrical–thermal analysis showed that the electric field was distributed unevenly across the multilayer insulation because the constituent materials had different temperature-dependent conductivities. The lower-conductivity fluoropolymer layers sustained higher electric stresses than the polyimide layers, and radial temperature gradients could produce field inversion within individual layers. For all proposed designs, the calculated maximum field in each layer remained below 80% of its adopted dielectric strength under the modeled conditions. These results supported the electrical feasibility of the candidate designs while highlighting the importance of conductivity compatibility between adjacent insulation layers [33].

These studies established candidate insulation constructions for the ±5 kV, 1 kA target. The next development was to combine the thermal model with the different current demands and pressures encountered during takeoff and cruise, rather than apply the takeoff current target at cruise pressure in every comparison.

### *G. Linking takeoff and cruise power flow to cable loading*

The initial low-pressure investigations combined the 1 kA design target with a cruise-representative pressure to examine a demanding thermal condition. The later work distinguished the two flight phases. The DCAS conference paper extended the aircraft power-flow analysis to cruise [34], and the KPEC conference paper used the resulting current requirement to compare bipolar cable designs at atmospheric and reduced pressures [35]. The journal study integrated the aircraft network analysis, contingency assessment, and coupled cable optimization [36]. The review therefore treats the journal results as the principal evidence and the conference papers as stages in developing the method.

The later network study retained the ±5 kV propulsion bus and fourteen motor–fan loads. It assigned approximately 25 MW to takeoff and 7.5 MW to cruise, so the modeled cruise propulsion demand was 30% of takeoff demand. This ratio was a loading assumption adopted for the investigated aircraft concept, informed by a representative thrust-power profile; it is not a universal ratio for all wide-body missions. The modified Z-bus solution was checked using the modified monotone-mapping method, and the network was examined under normal operation and single contingencies [34], [36]. The severe cruise contingency reported in the journal study involved the loss of busbar 6 and gave a maximum branch-current magnitude of approximately 363 A. The authors adopted 450 A as the cruise cable-design current to provide additional allowance for uncertainty, alongside the existing 1 kA takeoff design target. The adopted 450 A design current is approximately 24% above the calculated maximum branch current of 363 A, corresponding to approximately 81% utilization of the design current [35], [36]. The 30% total propulsion-power ratio should not be applied directly to the 1 kA takeoff target: the maximum branch current depends on the network, source allocation, and contingency, and the adopted cable-design currents include allowances.

The resulting design comparison used two operating points: 1000 A at 101.325 kPa for takeoff, and 450 A at 18.8 kPa for cruise. Takeoff imposes greater resistive heating, while cruise reduces buoyancy-driven cooling. Their effects oppose each other, so pressure alone cannot determine which case requires the larger conductor. The pressure values represent the local conditions assigned to the modeled installation; the low-pressure case applies to vented, unpressurized locations identified in Section I [35], [36].

## IV. Geometry Optimization of Bipolar MVDC Cables

Section III established insulation constructions, the coupled electrical–thermal–airflow model, and the takeoff and cruise current requirements. This section applies that framework to circular bipolar cables and alternative conductor geometries. The complementary geometry studies examined how conductor shape, the arrangement of the positive and negative poles, and their separation affect cooling, mass, and installation space. The 2023 CEIDP papers introduced coaxial and cuboid concepts [37], [38], which were developed further in the 2024 journal comparison of cylindrical, cuboid, and coaxial bipolar systems [39]. Rectangular poles were investigated in a 2024 TPEC conference paper [40] and a broader IEEE Access journal study [41]. The TPEC work was subsequently extended in a journal study of takeoff and cruise operating conditions [42]. The subsequent duct studies are discussed together in Section IV.F. The discussion below emphasizes the journal results and uses the conference studies to trace the development of the design approach.

### *A. From one cable to a bipolar circuit*

Two separately insulated poles in the same enclosure do not cool independently. Each pole blocks part of the other pole's radiative view of the enclosure and alters the nearby airflow. When their outer jackets touch, the facing surfaces cannot exchange heat with the surrounding air in the same way as exposed surfaces. Increasing their separation can improve heat rejection and allow smaller conductors for a specified current [39]–[42]. Section IV.F examines the additional installation space and duct material required to accommodate that separation.

A complementary bipolar ampacity study developed both a coupled electrical–thermal–airflow finite-element model and an analytical model incorporating conduction, convection, and radiation [43]. For circular poles at 18.8 kPa, it examined separation, horizontal and vertical placement, ambient temperature, duct size, and cable diameter. Under the investigated conditions, the horizontal arrangement could exceed the corresponding single-pole ampacity at sufficiently large separation because enhanced convection offset reduced radiative heat rejection. The vertical arrangement remained below the single-pole ampacity in the reported comparison. These results show why the current rating of a bipolar installation cannot be inferred directly from an isolated cable. Agreement between the analytical and finite-element predictions supports their use for preliminary sizing; it does not replace experimental validation of a manufactured installation.

Fig. 17 shows the principal geometries. The cylindrical arrangement uses two circular conductors; the cuboid arrangement uses two conductors with rounded square cross sections; and the rectangular arrangement uses flattened conductors with rounded corners. In these papers, "cuboid" describes the longitudinal cable body, whose transverse section is approximately square. A coaxial bipolar cable instead places one pole inside an annular return pole, separated by insulation. The concentric conductors carry equal-magnitude circuit current. With the inner pole at +5 kV and the outer pole at −5 kV, the dielectric between them supports 10 kV, compared with the nominal 5 kV between a separately screened pole and its grounded screen [39], [41].

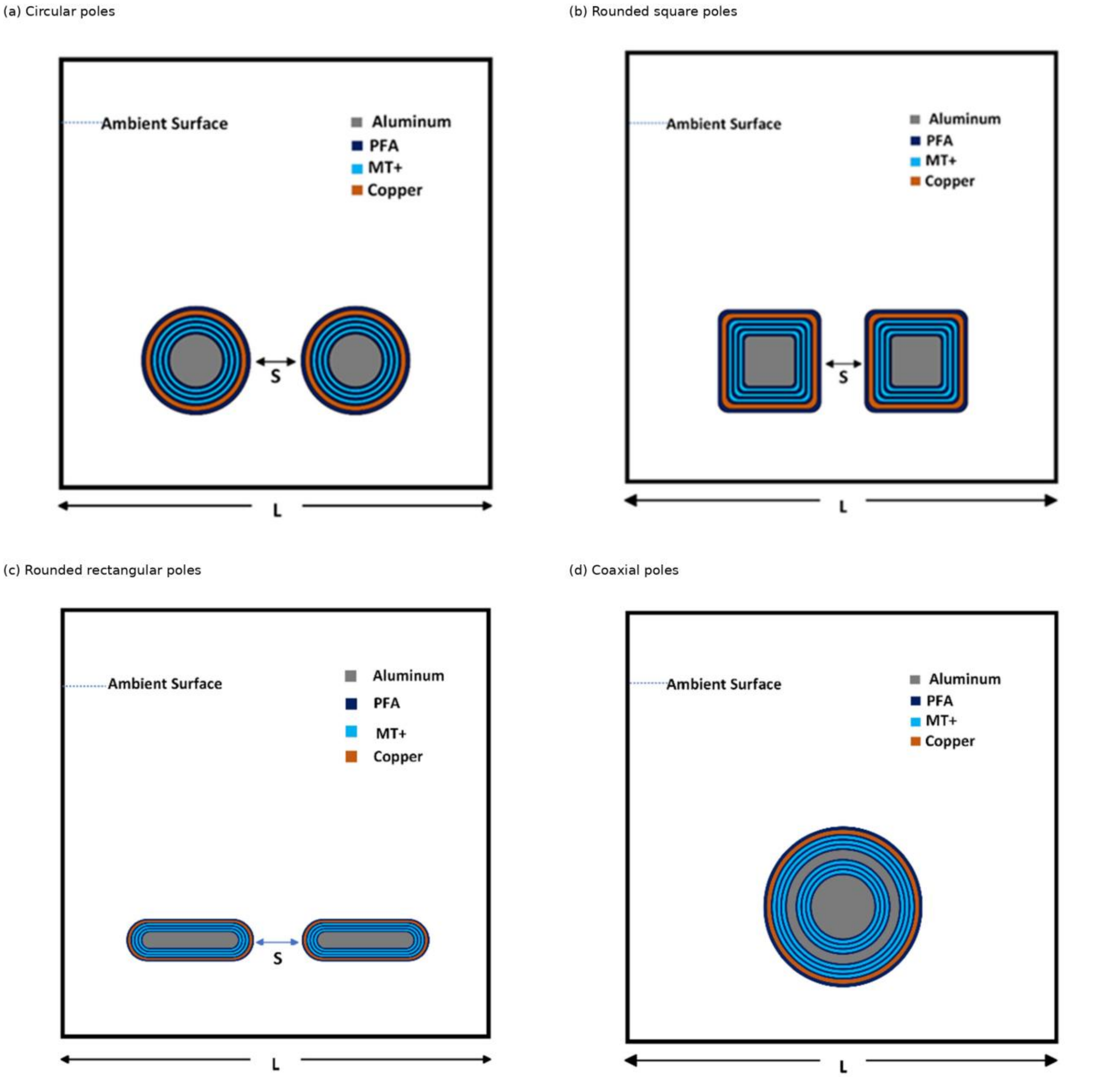

Fig. 17. Bipolar cable configurations studied in a square enclosure: (a) cylindrical, (b) cuboid with rounded square sections, (c) rounded rectangular, and (d) coaxial. S is the surface-to-surface separation of the two separate poles; L is the enclosure side length. Panels reproduced from [41].

The model retained heat conduction, natural convection, surface-to-surface radiation, and temperature-dependent DC conduction, as described in Section III.B. Pole separation S, conductor dimensions, and the selected cross-sectional shape became design variables. The separate-pole studies varied S from 0 to 2 in (0–50.8 mm); S = 0 denotes contact between outer cable surfaces, not contact between conductors. Rounded conductor and insulation corners avoided the artificial field singularities associated with ideally sharp corners. The governing equations and solution procedures are given in the source studies and are not repeated here.

For the 2024 high-temperature comparisons, each pole carried 1 kA at 18.8 kPa, with a 40 °C enclosure boundary, a 1 m square duct, and a nominal 1 in clearance above the duct floor. ARC-SC-T-MMEI insulation linked these studies to Section III, and a 260 °C temperature limit was used in the optimization [39], [41]. These are model conditions for comparing designs, rather than measured installation limits for an aircraft.

The objective also changed from the single-cable comparison in Section III.F. In the geometry studies [39], [41],

[42], $J$ denotes the product of cable mass per unit length and the total cross-sectional area of the insulated cables.

*B. Bipolar cable optimization under the two flight conditions*

The journal analysis used the circular bipolar ARC-SC-T-MMEI construction introduced earlier, with an aluminum conductor, alternating Kapton MT+/PFA insulation, a copper screen, and a PFA outer jacket. Both poles were included in the coupled model, so their mutual radiative blockage and airflow interaction were represented. Fig. 18 shows the geometry. A square aluminum enclosure of side length 250 mm surrounded a rounded rectangular open-boundary air domain extending 25.4 mm beyond the cable surfaces. The prescribed enclosure and ambient temperature was 40 °C. The inner air-domain outline is a computational boundary, not an additional physical duct wall; the separate duct-design investigations are discussed in Section IV.F [36]. The conductor diameter was varied separately for each operating point and each pole separation from 0 to 2 in (0–50.8 mm). The study adopted a 250 °C conductor-temperature limit and checked the insulation field against 80% of the adopted dielectric strength of each layer. Its cable-only design metric combined cable mass per unit length and cable cross-sectional area, rather than the mass–diameter metric used in Section III.F. The enclosure mass and the space between poles were not included in that cable-only metric. Simulations were run to a steady-state condition; the stated 30 h simulation interval was a numerical settling interval, not an aircraft takeoff or cruise duration [36].

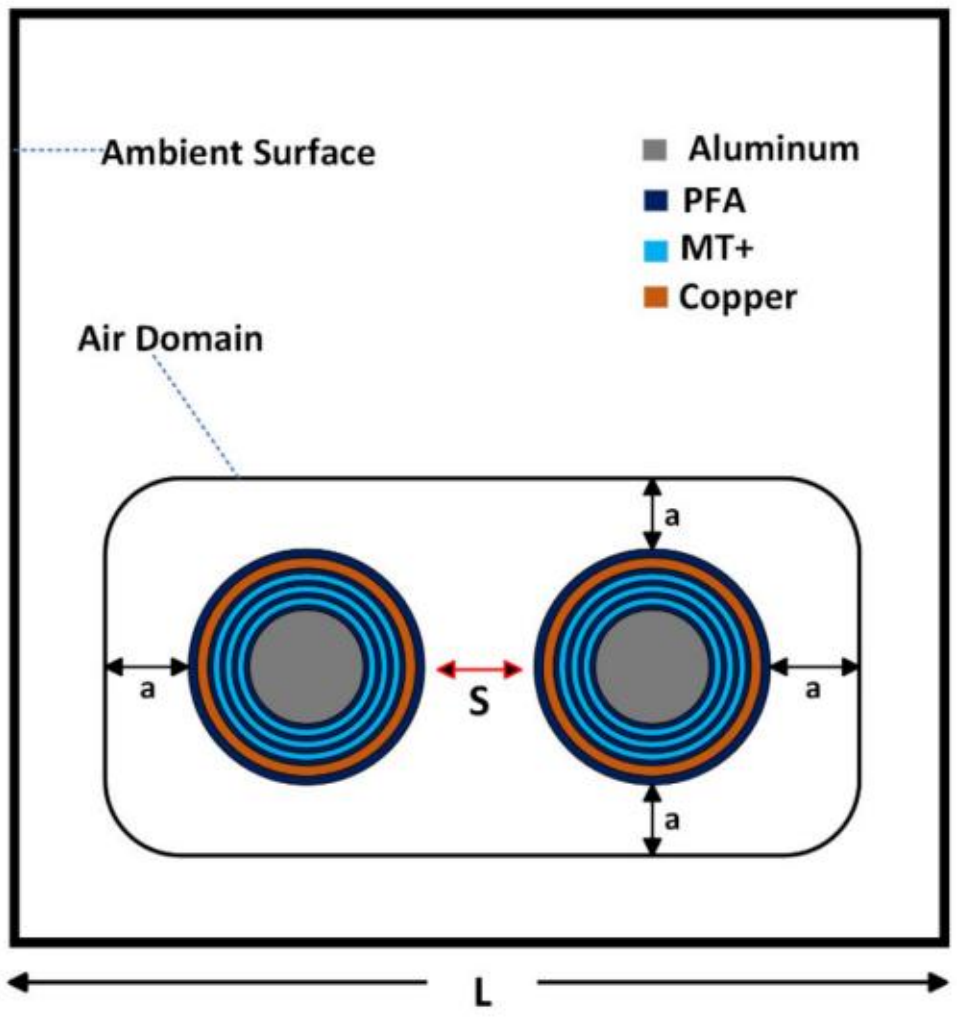

Fig. 18. Circular bipolar ARC-SC-T-MMEI cable model used for the takeoff and cruise comparison. S is the surface-to-surface pole separation, L is the enclosure side length, and a defines the air-domain clearance around the cables. The rounded inner outline is the open-boundary air domain [36].

At S = 2 in, the optimized takeoff construction had a conductor diameter of 14.06 mm, an insulated pole diameter of 14.85 mm, a combined cable cross section of 346.25 mm², and a total cable mass of approximately 0.988 kg/m. The independently optimized cruise construction had corresponding values of 8.85 mm, 9.64 mm, 145.87 mm², and approximately 0.428 kg/m. The takeoff construction therefore required about 2.37 times the cable cross section and 2.31 times the mass of the cruise construction at this spacing. Fig. 19 summarizes the dimensional and mass comparisons [36]. These are alternative sizing results for the two operating points, not dimensions that change during flight.

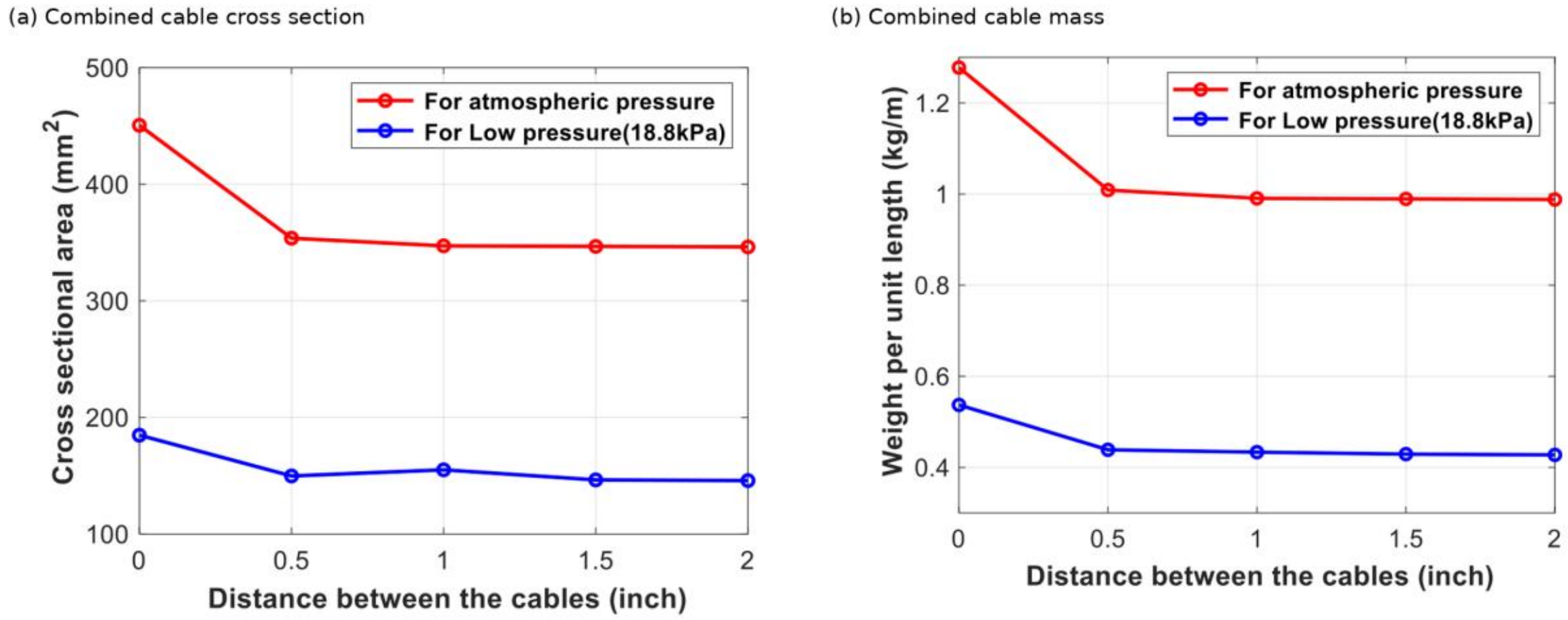

Fig. 19. Cable-only comparison of the separately optimized takeoff (1000 A, 101.325 kPa) and cruise (450 A, 18.8 kPa) constructions: (a) total cross-sectional area of both insulated poles and (b) total cable mass per unit length [36].

Separating the poles generally reduced the required mass and dimensions, with the lowest reported cable-only metric at 2 in within the investigated range. Most of the takeoff improvement occurred between touching poles and the first nonzero separation. At S = 0, the combined takeoff cross section was 450.6 mm² and the mass was approximately 1.278 kg/m; at S = 2 in, both were about 23% lower. This finding identifies the favorable end of the studied range for the cable-only objective. The installation penalty of a wider gap must be assessed separately, as discussed in Section IV.F [36].

For these two modeled operating points, the higher-current takeoff condition governed the required cable dimensions despite the weaker convection at cruise altitude. This was the main outcome of linking the network and cable models [35], [36]. The initial 1 kA low-pressure condition remains useful as a demanding screening case, but it should be distinguished from the phase-specific design comparison. The installed construction must meet both operating points; separate sizing calculations identify the candidate governing case, while checking one fixed cable over its load history is the subsequent verification step. These results also provide the loading basis for the subsequent comparisons of alternative conductor geometries and the duct studies in Section IV.F.

### *C. Rounded square and coaxial alternatives*

The initial cuboid conference study used a conventional NL-EPR-based construction to examine the benefit of reshaping a cable while maintaining comparable material quantity and maximum electric stress. At 18.8 kPa and 65 °C ambient, its rounded square cable had approximately 3% and 4.6% greater permissible current than the cylindrical comparator in 250 and 1000 mm ducts, respectively [38]. This established the potential thermal benefit of a larger exposed perimeter, but the gain depended on the surrounding radiative environment. A larger perimeter does not by itself guarantee a proportional increase in convection, because shape also changes the airflow.

The journal extension compared complete bipolar systems using the high-temperature multilayer construction. At S = 2 in, the cuboid pair weighed 1.4901 kg/m and occupied 524.53 mm² of combined cable cross section, compared with 1.5320 kg/m and 539.18 mm² for the cylindrical pair. At S = 2 in, the cuboid pair had a lower cable mass-and-area objective, J, than the cylindrical pair [39]. At zero separation, however, its $J$ was approximately 2.82% higher. Thus, the rounded square configuration improved the combined cable mass-and-area objective over much of the spacing range, but did not outperform the circular pair under every arrangement [39].

The coaxial conference study found that concentrating both poles inside a common outer surface could reduce heat rejection relative to two exposed cables [37]. The journal comparison reached the same practical conclusion for the solid-inner-conductor coaxial design: its mass was 1.8329 kg/m and its overall diameter was 29.052 mm, making it less favorable in the cable mass-and-area comparison than well-separated cylindrical or cuboid poles [39]. The annular return conductor and the larger voltage across the inter-pole dielectric must therefore be included when judging whether concentric packaging is advantageous.

A hollow inner conductor provides a way to enlarge the outside heat-rejection surface without filling the whole interior with metal (Fig. 20). In the journal study, a 50 mm hollow diameter reduced coaxial cable mass to approximately 13.8% below the cylindrical pair and 11.4% below the cuboid pair at S = 2 in. This mass saving came with an occupied cable cross section approximately 4.5–4.6 times as large and a reported J approximately 3.9–4.1 times as large. Hollow coaxial construction was therefore a mass-versus-space tradeoff, rather than the preferred design under the combined objective [39].

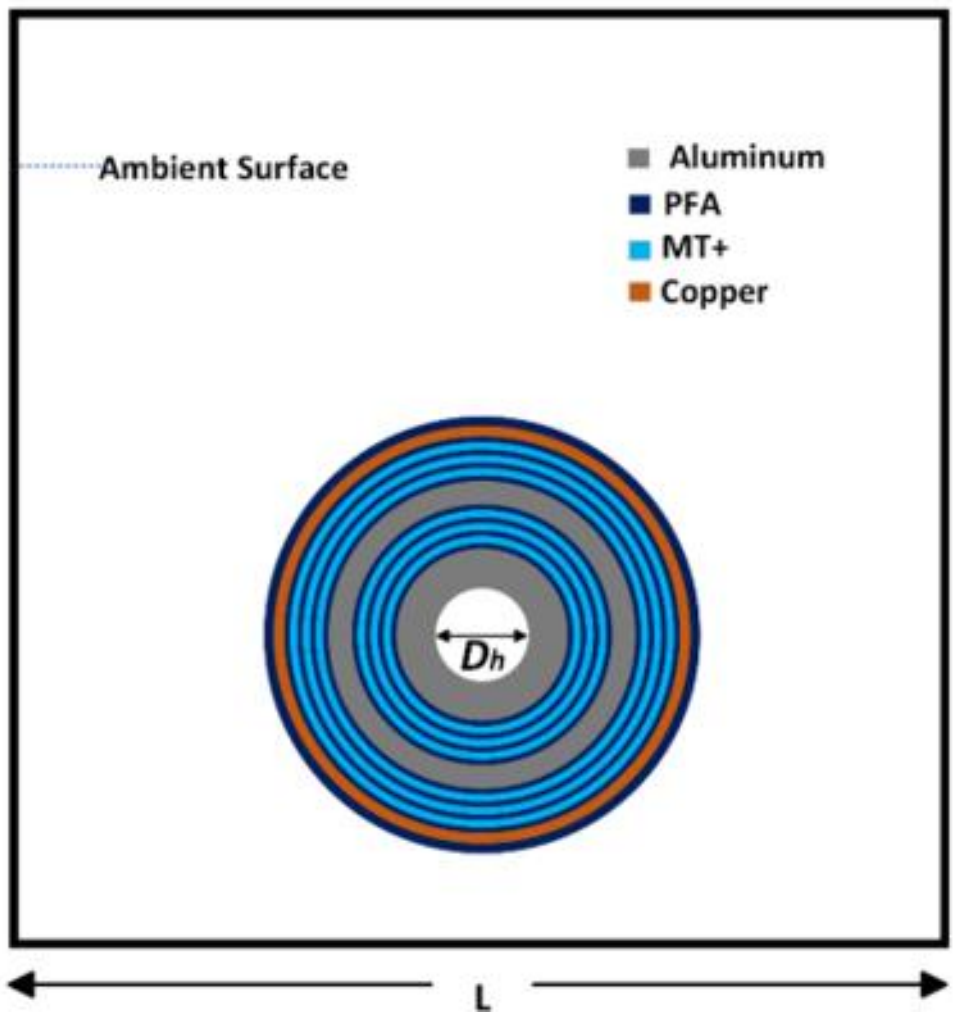

Fig. 20. Coaxial bipolar configuration with a hollow inner conductor. The hollow diameter is an additional design variable; setting it to zero gives the solid-inner-conductor case [39].

### *D. Rectangular poles and separation as design variables*

The rectangular conference study compared NL-EPR-based cable shapes at the same maximum temperature and comparable electrical stress. At S = 2 in and 18.8 kPa, its rectangular pair carried approximately 1230 A, about 8.3% more than the cylindrical pair and 5.6% more than the cuboid pair [40]. The IEEE Access journal investigation then compared optimized rectangular, cuboid, cylindrical, and coaxial cables using the high-temperature multilayer construction at a fixed current of 1 kA [41]. These are different comparisons: the conference result concerns permissible current for the selected constructions, whereas the journal result concerns dimensions and mass needed to carry a prescribed current.

The journal rectangular pair had the smallest reported J throughout the studied separation range. At S = 2 in, its combined mass was 1.3548 kg/m and its combined cable cross section was 468.97 mm². Its mass was approximately 11.5% below the cylindrical pair, 9.1% below the cuboid pair, and 26.1% below the solid-inner-conductor coaxial cable. The corresponding reported $J$ was 635.36 g·mm, approximately 22.6%, 18.7%, and 47.7% below those respective comparators [41]. Table IV summarizes the cable dimensions and masses; Fig. 21 places the rectangular result alongside the hollow-coaxial tradeoff.

TABLE IV. CABLE-ONLY COMPARISON AT 1 kA, 18.8 kPa AND 40 °C [39], [41]

| Configuration | Separation S | Overall dimensions of one cable (mm) | Combined cable area (mm²) | Cable mass (kg/m) |
|---|---|---|---|---|
| Cylindrical pair | 2 in | Diameter 18.527 | 539.18 | 1.5320 |
| Cuboid pair | 2 in | Side 16.28 | 524.53 | 1.4901 |
| Rectangular pair | 2 in | 32.84 × 7.20 | 468.97 | 1.3548 |
| Coaxial, solid inner | Concentric | Diameter 29.052 | 662.86 | 1.8329 |

For the rectangular pair, the reported J fell from 821.98 g·mm at S = 0 to 635.36 g·mm at S = 2 in, a reduction of approximately 22.7%. Increased separation improved radiative exchange with the enclosure and exposed the facing surfaces to airflow. The cooling gain allowed smaller conductors and lower mass at the prescribed current, although the rate of improvement diminished as separation increased [41]. The 2 in result is the lowest value within the investigated range and objective; it is not a universal optimum for an installation that must also accommodate the gap and duct.

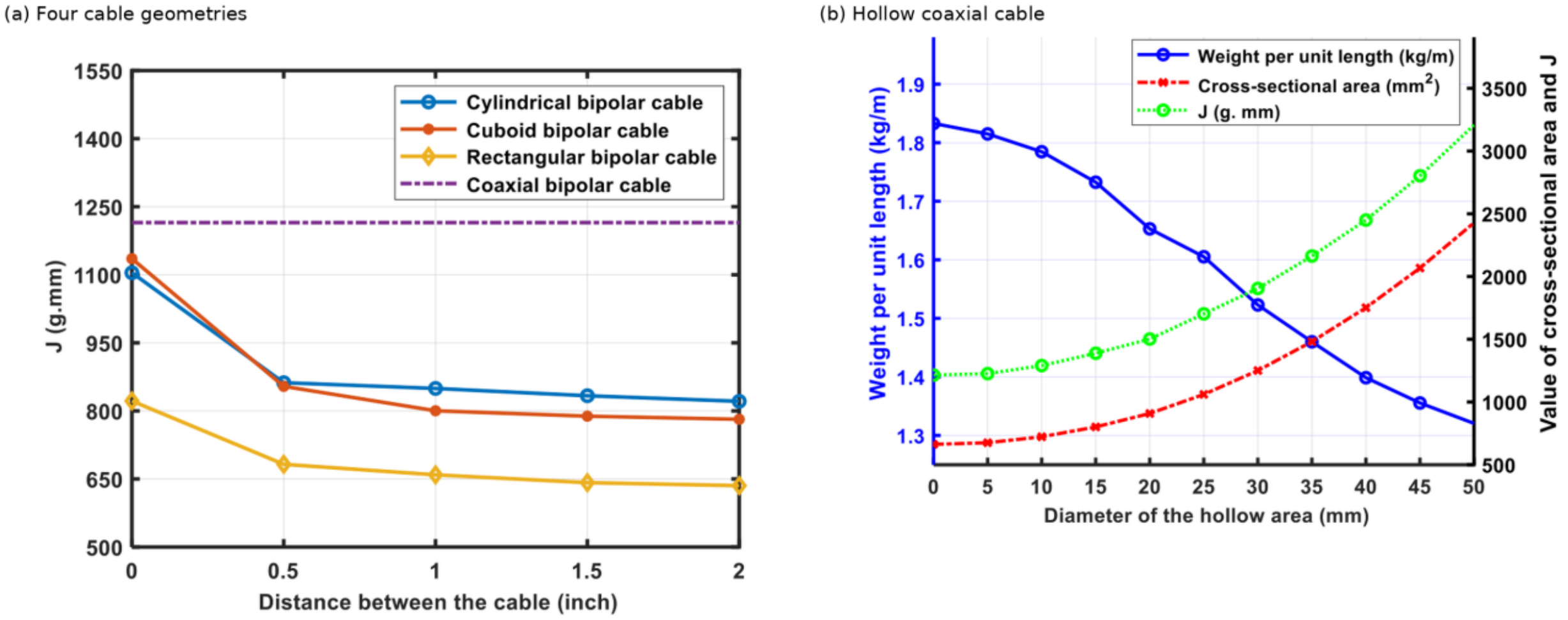

Fig. 21. Selected results from the high-temperature geometry studies: (a) reported cable-only J for cylindrical, cuboid, rectangular, and solid-inner-conductor coaxial systems as pole spacing varies [41]; (b) mass, cable cross section, and J as hollow diameter varies in a coaxial system [39]. These calculations use 1 kA, 18.8 kPa, and 40 °C ambient. Panel (b) retains the source's separate vertical scales.

The electrical analyses checked the maximum field in each insulation layer against the adopted dielectric-strength criterion. The coaxial inter-pole insulation experienced greater stress because it supported 10 kV. For the rounded separate-pole designs, the calculated fields remained within the study criterion [39], [41].

### *E. Takeoff and cruise conditions can change the preferred geometry*

The journal extension of the rectangular TPEC study evaluated two operating points: 1000 A at 101.325 kPa for takeoff and 450 A at 18.8 kPa for cruise [42]. Unlike the earlier 1 kA low-pressure multilayer comparison, it used NL-EPR insulation and a PVC jacket, a 105 °C temperature limit, and a 250 mm square duct. The conductor sizes were optimized separately for the two operating points. The smaller dimensions obtained for cruise are therefore alternative sizing results, not a change in the physical size or mass of one installed cable during flight. The power-flow basis for these two current targets is explained in Section III.G; this study extends the comparison to alternative conductor shapes and a different insulation construction.

At S = 2 in in the takeoff case, the rectangular pair had a mass of approximately 6.08 kg/m, about 2% below the cylindrical pair and 1.18% below the cuboid pair. Its greater external perimeter, however, required more insulation and jacket material at the selected layer thicknesses and resulted in a larger overall cable cross section. The cylindrical pair had the smallest cable area, approximately 1342.6 mm², and the smallest J. The rectangular J was approximately 12% higher [42]. Improved heat rejection and slightly lower mass did not therefore guarantee the best combined mass-and-area objective.

The cruise comparison showed a similar distinction. At S = 2 in the masses were close to 2.6 kg/m, while the cylindrical pair had the smallest cross section, approximately 736.92 mm². The published J curves again placed the cylindrical pair lowest (Fig. 22). These results do not contradict the earlier rectangular advantage: the insulation construction, temperature limit, enclosure size, and operating current had changed [41], [42]. Geometry selection must consequently be tied to the intended materials and installation rather than inferred from conductor shape alone.

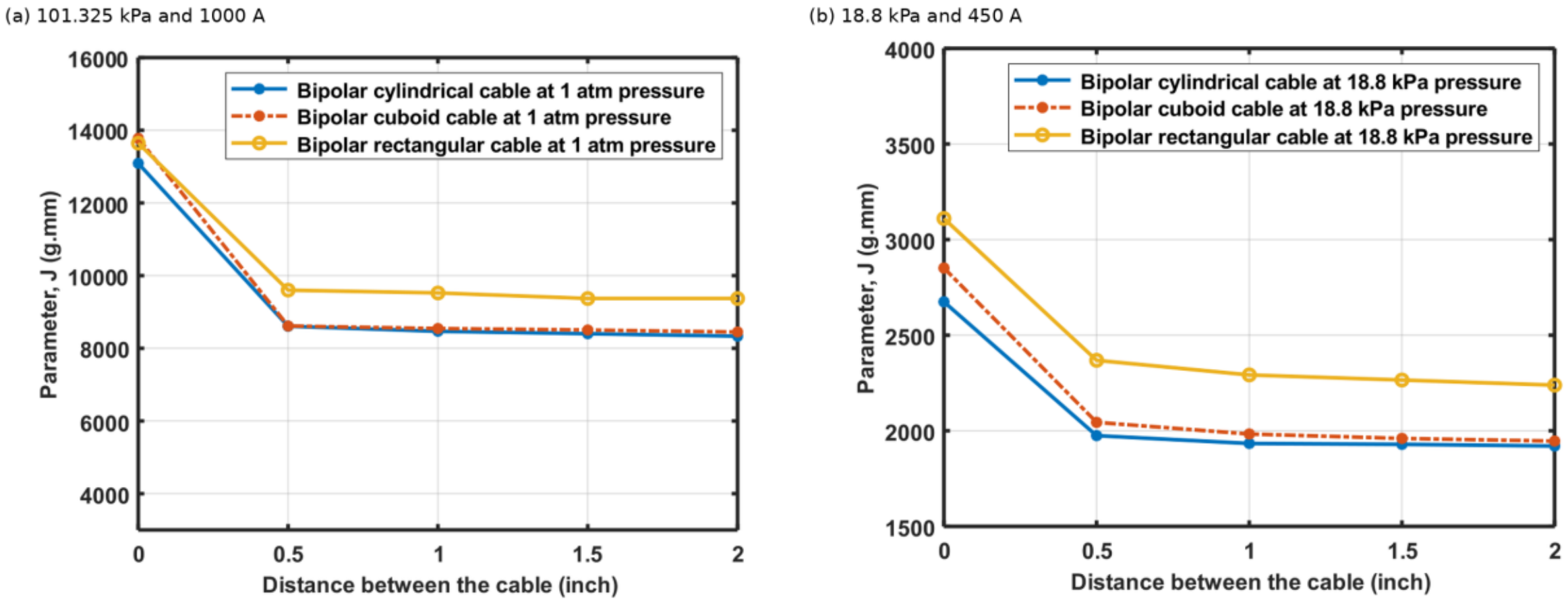

Fig. 22. Cable-only J for the NL-EPR/PVC designs under (a) takeoff and (b) cruise sizing conditions. Panels reproduced from [42]. The two panels represent separately optimized dimensions under different current and pressure conditions.

For the two operating points examined, the 1000 A takeoff case required the larger dimensions despite its higher pressure, because its current-related heating outweighed the cooling disadvantage of the lower-pressure 450 A case [42]. This finding establishes which of those two steady-state cases governed the reported sizing. A final aircraft cable must be checked as one fixed construction over the required mission and contingency envelope, including the duration of each load and its local cooling conditions. Reduced cruise current is an operating-load change; it does not, by itself, determine the ampacity of a cable.

### *F. Duct Design and Bipolar Cable Installation*

The duct is part of the thermal and structural design of the bipolar cable installation. Its shape determines the available airflow paths and radiative exchange surfaces, while its wall material and dimensions contribute additional mass and occupied space. Our duct investigations addressed two related questions: how circular and rounded square cables compare within a rounded rectangular duct [44], and how the duct shape itself changes the performance of circular bipolar cables. The latter comparison was presented at KPEC [45] and broadened in the journal investigation to include vertical pole placement and a central aluminum barrier [46]. This section brings these studies together, emphasizing the journal findings while retaining the conference studies as steps in the development of the installation design.

#### *F.1. From cable mass to the complete installation*

The coupled model described in Section III.B was applied to both poles, the enclosed air, and the aluminum duct. Conduction through the cable layers was coupled to buoyancy-driven airflow and surface-to-surface radiation across the air gap. The studies relied on natural convection; forced airflow was not imposed. The ARC-SC-T-MMEI construction linked the duct investigations to the earlier insulation studies. The governing equations and material descriptions are available in [44]–[46] and are not repeated here.

The common design conditions included a 1 mm aluminum wall, a cable-to-wall clearance of 25.4 mm, a 40 °C ambient temperature, and a surface-to-surface pole separation S varied from 0 to 50.8 mm. The clearance is the value adopted in these studies. Zero pole separation means that the outer cable surfaces touch. Two sizing conditions were considered: 1000 A at 101.325 kPa and 450 A at 18.8 kPa, representing takeoff and cruise, respectively. Cable dimensions were optimized separately for each condition. The comparison metric combined total cable-and-duct mass per unit length with the overall installation cross-sectional area, including the enclosed space. It therefore differs from the cable-only metric in Sections IV.A–IV.E [44]–[46]. The aircraft-level basis for the two current targets is discussed in Section III.G.

The SDEMPED study compared cylindrical and cuboid poles within rounded rectangular ducts [44]. Fig. 23 shows the two arrangements. The duct dimensions adjusted with the cable dimensions and pole separation to maintain the adopted cable-to-wall clearance. Consequently, a cooling benefit that reduced cable mass could be offset by the mass and space needed for a larger duct.

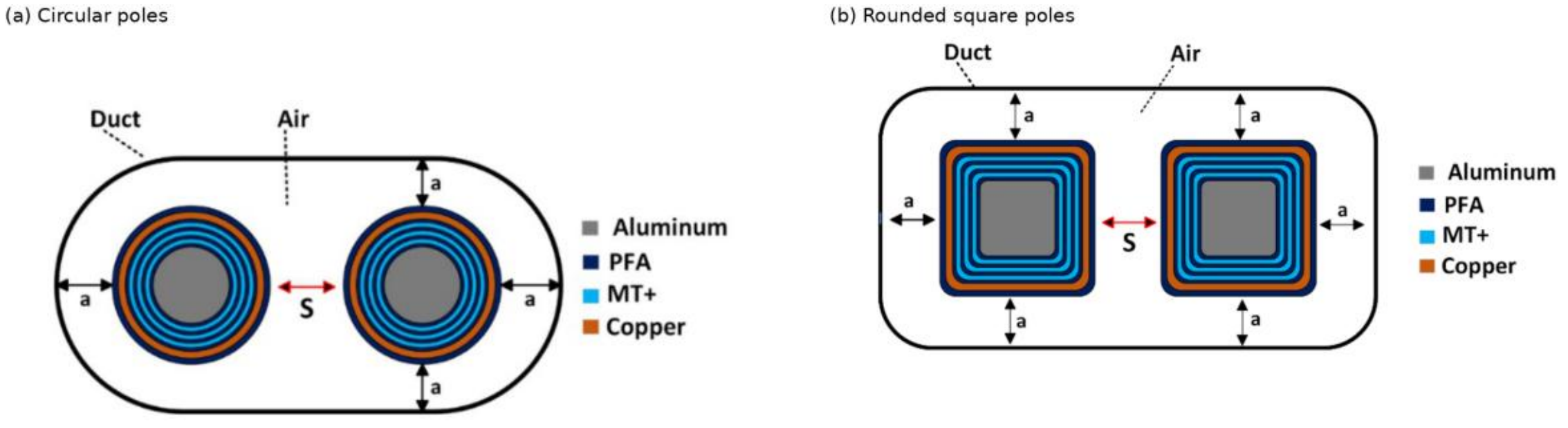

Fig. 23. Cylindrical and rounded square bipolar cables inside rounded rectangular ducts. The minimum cable-to-duct clearance a is 1 in, and the surface-to-surface pole separation S is varied. Panels reproduced from [44].

Under the 450 A, 18.8 kPa cruise sizing condition, the cuboid installation at S = 0 had a combined cable-and-duct mass of 1.4138 kg/m, compared with approximately 1.5475 kg/m for the cylindrical installation, an 8.6% reduction. Its overall cross section was also slightly smaller: 4467.5 mm² compared with 4483.0 mm². Although separating the poles reduced the cable mass needed for cooling, it increased the duct dimensions and mass. For the cuboid cruise design, total mass rose to 1.4671 kg/m at S = 2 in, while the overall cross section increased to 7390.0 mm² [44]. For the cuboid cruise case, zero separation gave the lowest combined mass-and-area objective over the studied range. Thus, the favorable spacing for cable-only mass cannot be transferred directly to the complete installation.

The takeoff comparison did not produce a consistent cuboid advantage. At S = 0, total masses were nearly equal: 2.8957 kg/m for the cuboid installation and 2.9062 kg/m for the cylindrical installation. At S = 0.5 in, the cuboid installation was heavier, 2.8587 kg/m compared with 2.6147 kg/m for the cylindrical one [44]. These results show why the conductor shape, insulation, duct dimensions, and operating condition must be evaluated together.

*F.2. Duct shape and pole orientation*

The KPEC study compared rounded rectangular and cylindrical ducts containing circular poles arranged side by side [45]. The journal investigation added a cylindrical duct with vertically stacked poles [46]. Fig. 24 shows these three layouts and the barrier variant discussed in Section IV.F.3. Horizontal and vertical placement describe the arrangement in the transverse section relative to gravity. A vertical arrangement places one pole above the other and changes the interaction of their buoyant plumes; it does not refer to the direction of cable routing along the aircraft.

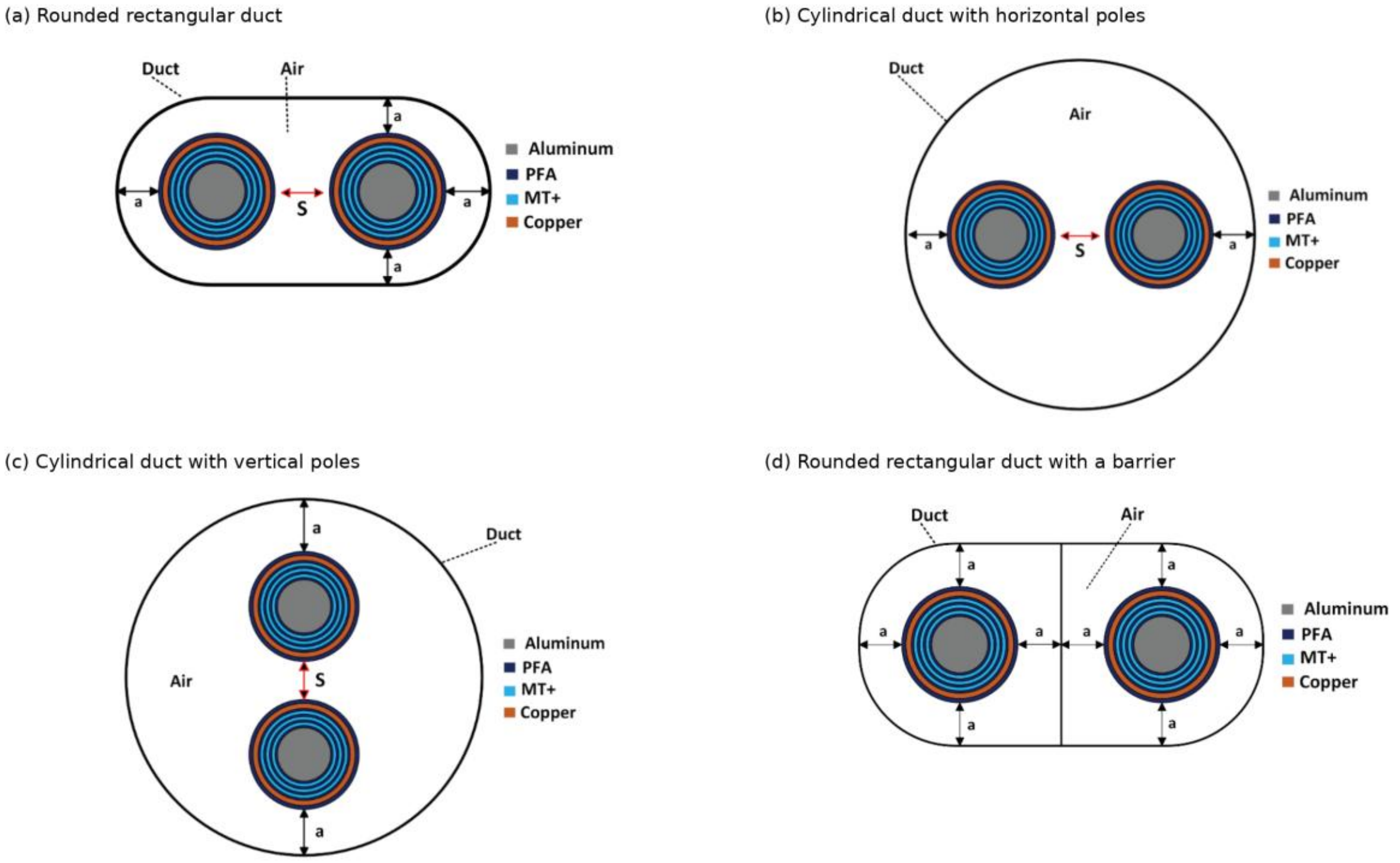

Fig. 24. Duct configurations for circular bipolar cables: (a) rounded rectangular, (b) cylindrical with horizontal pole placement, (c) cylindrical with vertical pole placement, and (d) rounded rectangular with a central aluminum barrier. The cable-to-wall clearance is a; S denotes pole separation in the unpartitioned arrangements. Panels reproduced from [46].

In the reported takeoff sizing results, cylindrical ducts had the lower combined mass and mass-and-area metric at zero pole separation. The horizontal and vertical cylindrical installations weighed 2.2695 and 2.2326 kg/m, respectively, compared with 2.9062 kg/m for the rounded rectangular installation. The rounded rectangular duct nevertheless occupied slightly less cross-sectional area: 6068.4 mm² compared with 6328.9 and 6269.8 mm². Thus, minimum mass and minimum installation area did not identify the same arrangement [46].

As separation increased, the rounded rectangular duct could grow mainly in width, whereas a cylindrical duct required a larger diameter and consequently expanded in both transverse directions. At S = 50.8 mm, the rounded rectangular installation occupied 9171.2 mm², compared with 14 886 and 14 834 mm² for the horizontal and vertical cylindrical installations. Its reported combined mass-and-area metric was approximately 33% lower than those of the cylindrical alternatives. This percentage uses the cylindrical installations as the comparison baseline. Fig. 25 shows the crossover: rounded rectangular ducts were favorable at wider spacing, while cylindrical ducts were favorable in the most closely packed arrangements [46].

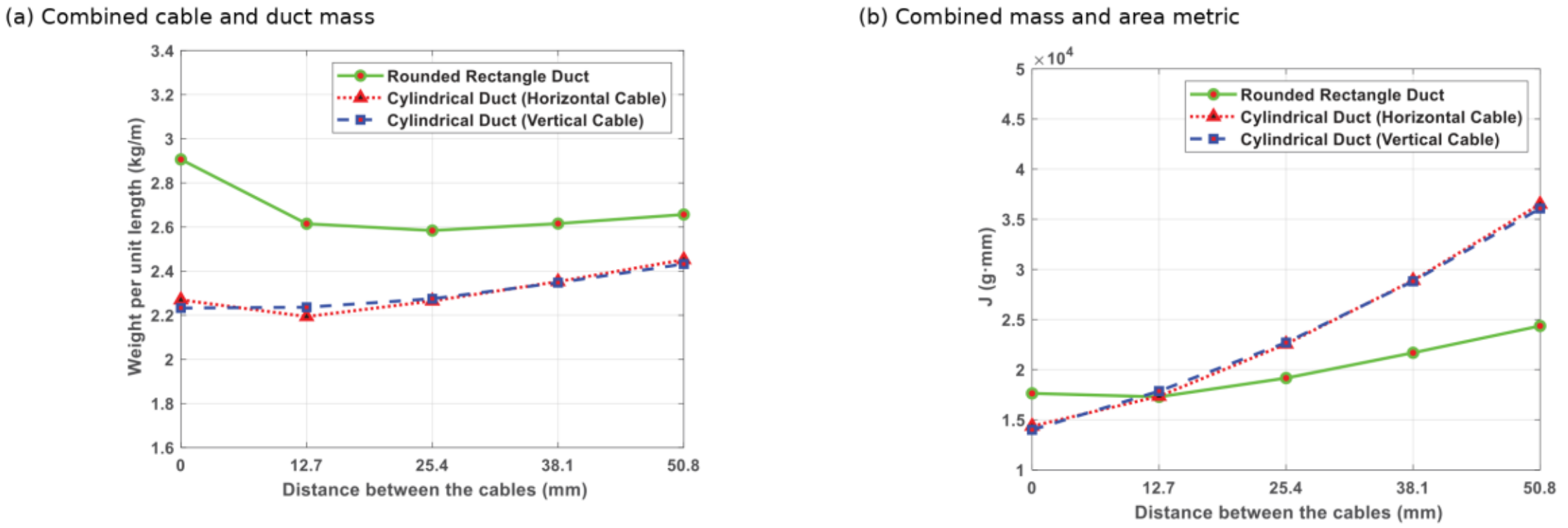

Fig. 25. Selected journal results for the 1000 A, 101.325 kPa sizing condition: (a) total cable-and-duct mass per unit length and (b) the product of that mass per unit length and the overall installation cross-sectional area. Panels reproduced from [46]. Each spacing corresponds to a separately sized cable and duct.

The heat-transfer results explain why the structural ranking cannot be inferred from one cooling mechanism alone. At wider spacing, cylindrical ducts provided greater radiative heat transfer, while the rounded rectangular duct provided greater convective heat transfer in the reported atmospheric-pressure comparison. The journal airflow maps showed different buoyant circulation patterns in the two shapes [46]. Pole orientation also changed the thermal interaction, although its influence on the combined mass-and-area result was smaller than the difference between duct shapes at large spacing. A higher radiative or convective contribution by itself therefore does not establish the lightest or most compact complete installation.

For each layout, the reported 1000 A takeoff sizing case required larger cable dimensions and total mass than the 450 A cruise sizing case, despite the lower air pressure during cruise [45], [46]. This is consistent with Section IV.E: the higher current can govern the thermal design even when convection is more effective. These results compare separately optimized constructions. One installed cable and duct must ultimately satisfy both operating conditions and the relevant mission durations.

*F.3. Central barrier in a rounded rectangular duct*

The journal study also introduced a central aluminum barrier with the same 1 mm thickness as the duct wall (Fig. 24(d)). Varying its height changed the surfaces available for radiation and the air circulation between the poles, while adding structural material. The calculated cable dimensions decreased slightly as barrier height increased, but this benefit was largely offset by the additional barrier mass [46].

Over the reported height range of approximately 44.5–69.5 mm, the overall installation cross section decreased from about 9102.6 to 9039 mm², a reduction of approximately 0.7%. The combined mass-and-area metric remained close to $2.44 \times 10^4$ g·mm and was lowest at the smallest investigated barrier height; the change across the studied range was only about 0.2%. Radiative and convective heat-transfer contributions also changed relatively little [46]. The barrier therefore offered only a small dimensional benefit in this comparison. Its addition should not be described as a substantial thermal improvement or an established optimum outside the investigated height range.

The electrical analysis checked the cable insulation against the adopted maximum-field criterion [46]. This supports the modeled dielectric design, but does not establish that the central barrier provides fault containment or arc isolation; those functions would require separate evaluation.

### *F.4. Implications for cable installation design*

The duct studies establish that pole spacing and conductor geometry must be optimized with the enclosure when the objective is aircraft installation mass and space. Separation can improve cable cooling, yet increase the duct dimensions enough to worsen the combined result. Rounded square poles reduced mass in some rounded-duct comparisons [44], and rounded rectangular ducts limited the growth in occupied area at wider pole spacing [45], [46]. Cylindrical ducts remained competitive at small separations. These are conditional design rankings tied to the specified materials, clearances, pressure, current, and thermal boundaries.

The insulation, geometry, and duct studies provide a numerical basis for selecting the cable construction and its installation. The following section examines whether imposed airflow offers further reductions in cable mass and dimensions.

## *G. Forced Air Cooling of Bipolar MVDC Cables*

The preceding designs relied on buoyancy-driven natural convection and radiation to reject cable heat. We next investigated imposed airflow as an additional design variable: stronger convective cooling can lower the temperature of a fixed cable, or permit a smaller conductor at the same current and temperature limit. The 2024 DCAS conference paper introduced this comparison for circular bipolar ARC-SC-T-MMEI cables at 18.8 kPa [47]. The subsequent journal study extended the analysis to both 18.8 and 101.325 kPa and examined the associated thermal and electric-field behavior [48]. The main results below are taken from the journal study, with [47] documenting the initial low-pressure investigation.

### *G.1. Model and comparison conditions*

The study retained the screened multilayer construction described in Section III and the circular bipolar arrangement introduced in Section IV.A. The radiation enclosure was a square aluminum duct of side length 1000 mm and wall thickness 1 mm, with the poles positioned 25.4 mm above its floor and separated by 50.8 mm. The forced-flow analysis considered a 20 m duct with one inlet and one outlet. The prescribed ambient and enclosure temperature was 40 °C. Air velocity was varied from 1 to 10 m/s in the journal analysis; the initial conference study considered 5–10 m/s at low pressure [47], [48]. The governing heat-transfer and airflow equations and boundary conditions are given in those papers and are not repeated here.

For each pressure and cooling condition, conductor diameter was adjusted to carry 1000 A while reaching the adopted 260 °C conductor-temperature limit. The voltage remained ±5 kV, and the comparison used cable mass per unit length, combined cable cross-sectional area, and their product J. This is a fixed-current comparison of cooling strategies: 1000 A was applied at both pressures. It therefore differs from the phase-specific comparison in Sections III.G and IV.B, where the cruise design current was 450 A. The low-pressure results here should not be presented as a new estimate of cruise current demand [48].

At a given temperature and characteristic size, lower air pressure reduces air density and the Reynolds number associated with a prescribed velocity. Consequently, the same air speed does not provide the same cooling at the two pressures. The simulations addressed this pressure dependence by comparing each forced-flow case with its corresponding natural-convection baseline [48]. The reported velocity thresholds are specific to the modeled geometry, airflow treatment, and thermal boundaries.

### *G.2. Predicted mass and size benefits*

At 18.8 kPa, the reported natural-convection baseline had a cable mass of 1.5342 kg/m and J of approximately 833.47 g·mm. Forced convection produced a lower J than this baseline only when the imposed velocity exceeded approximately 7 m/s. At 8 m/s, the reported reductions were about 5.2% in J and 2.6% in cable mass. At 10 m/s, J fell to approximately 713.03 g·mm and mass to 1.4203 kg/m, reductions of approximately 14.45% and 7.42%, respectively. The reported combined cable cross section decreased by about 7.59% at 10 m/s. Thus, the reduction in J should not be described as an equal percentage reduction in cable mass or an experimentally demonstrated increase in power density [47], [48].

At 101.325 kPa, the natural-convection baseline was lighter and smaller, with a mass of 1.2312 kg/m and J of approximately 534.05 g·mm. The forced-convection J became lower than the natural-convection value above approximately 3 m/s. At 4 m/s, the reported J was 458.38 g·mm, about 14.2% below the baseline. At 10 m/s, the reported mass was 0.8303 kg/m and J was 240.45 g·mm, approximately 32.6% and 55.0% below their respective natural-convection values [48]. Fig. 26 compares the two pressures. These curves represent separately optimized cable dimensions, rather than the temperature response of one fixed cable as airflow changes.

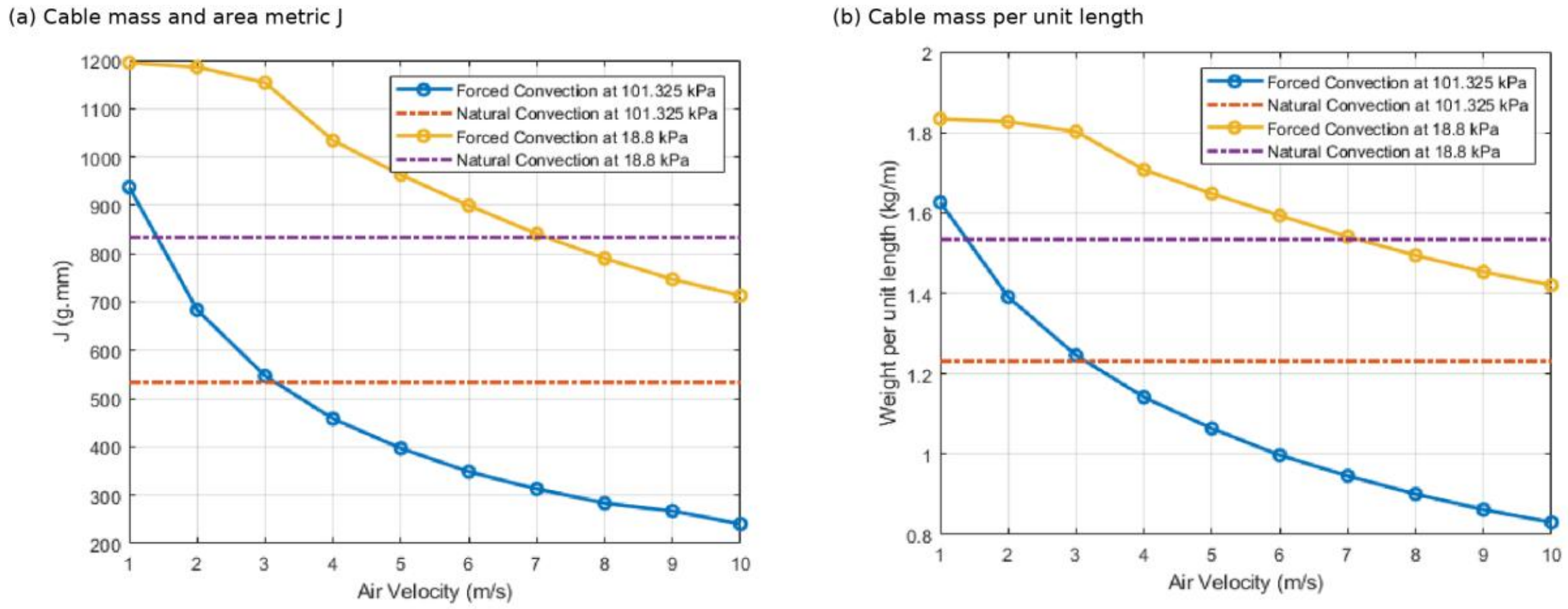

Fig. 26. Comparison of forced convection with the corresponding natural-convection baselines at 101.325 and 18.8 kPa: (a) the product of cable mass per unit length and combined cable cross-sectional area, J, and (b) cable mass per unit length. Current is 1000 A in every case, and each point is optimized against the same 260 °C conductor-temperature limit. Panels reproduced from [48].

The atmospheric-pressure results show a greater benefit at the same imposed air speed. At lower velocities, however, the forced-flow cases in this model required larger cable dimensions than the corresponding natural-convection solutions. The findings therefore support a conditional choice of forced-air cooling, rather than the general claim that any imposed airflow improves the optimized design [48].

### *G.3. Electrical behavior and installation implications*

The journal analysis also examined DC field redistribution through the multilayer insulation. The optimized cases had similar temperature profiles, with a reported radial temperature drop of about 5 °C, and calculated peak insulation fields close to 73 MV/m at both pressures. The paper explicitly treats the polymer conductivity coefficients as phenomenological because validated temperature- and field-dependent data for the selected MT+/PFA layers were unavailable over the full operating range. The electrical results should therefore be used as qualitative comparisons of field redistribution, rather than validated predictions of dielectric endurance or partial-discharge behavior [48].

The reported reductions concern the cables alone. Fans or blowers, their power consumption, the pressure drop along the duct, and the mass of airflow equipment were not included in *J*. A practical installation decision must compare those additional demands with the cable mass saved and establish the cable loading permitted if airflow is reduced or lost. In particular, the crossover velocities are not aircraft-wide design limits. They identify where forced cooling became favorable within this numerical comparison [48].

The aircraft studies establish candidate insulation constructions, conductor geometries, and cooling arrangements. A related line of work extended the same design framework to lunar power transmission, where the surrounding environment changes the dominant heat-transfer paths, as discussed next.

## *H. Extension to Lunar Power Transmission*

The aircraft cable studies also provided a starting point for bipolar MVDC power transmission on the Moon. This related application shares the need to minimize conductor and insulation mass while managing electrical stress, but its vacuum environment changes the thermal design. The TPEC conference study examined buried coaxial cables at 500 kW and 1 MW with ±5 and ±10 kV voltage levels [49]. The CEIDP paper investigated a 2 MW, ±10 kV buried coaxial design and the effect of a longer thermal simulation [50]. The journal investigation broadened the comparison to surface and subsurface installations, copper and aluminum conductors, and power levels of 100 kW, 500 kW, and 1 MW at both voltage levels [51]. We emphasize the journal comparison here and use the conference results to show how the approach developed.

An additional Aerospace journal study examined buried bipolar coaxial cables at 200 kW, 1 MW, and 2 MW, comparing ±5 and ±10 kV and copper and aluminum conductors [52]. Its coupled electrothermal model used Kapton MT+ insulation and an installation approximately 30 cm below the lunar surface. Aluminum offered a lower-mass solution in the reported medium- and high-power comparisons, while the higher voltage reduced the current required for a given transmitted power. This study adds a buried-cable design comparison to the progression from the conference investigations [49], [50] to the broader surface/subsurface journal analysis [51]. Its findings remain conditional on the adopted regolith and thermal boundaries and do not provide a direct aircraft altitude derating factor.

### *H.1. Changes in geometry and thermal boundaries*

In lunar vacuum, natural and forced convection to ambient air are unavailable. The buried-cable model therefore transferred heat by conduction into the surrounding regolith, whereas the surface model included conduction to the ground and radiation, including absorbed sunlight and radiative heat rejection. Fig. 27 shows the installation geometries and coaxial cross section. These models extend the earlier coupled electrical–thermal framework by replacing the aircraft air-domain cooling boundary with regolith conduction and, for the surface cable, radiative exchange [49]–[51].

The journal model placed the buried cable 30 cm below the surface, prescribed a 250 K (−23.15 °C) boundary 5 cm below the cable, and used a constant regolith thermal conductivity of 0.01 W/(m·K). The upper ground boundary alternated between prescribed hot and cold values of 384 and 102 K in a simplified cyclic simulation extending to 2000 h. For the surface installation, the model included a varying solar heat input with a peak near 1361 W/m². These are the source study's thermal boundaries; the 250 K subsurface value and the imposed cycle should not be treated as universal conditions at every lunar location. The low regolith conductivity makes the simulation duration and the placement of thermal boundaries important to predicted cable temperature [51].

The coaxial cable used a central positive conductor and an annular negative return conductor with equal conducting areas. Alternating Kapton MT+ and PFA films formed the multilayer insulation, with a copper screen and a PFA outer jacket. The adopted film gauges were 1.5 mil for MT+ and 0.5 mil for PFA; layer counts and conductor dimensions varied with voltage, power, and installation. At ±5 kV, the inter-pole dielectric supports 10 kV; at ±10 kV, it supports 20 kV. For example, 1 MW corresponds to 100 A at ±5 kV or 50 A at ±10 kV, while the 2 MW conference design at ±10 kV carries 100 A. These lunar cases therefore have different current requirements from the 1 kA aircraft screening studies [49]–[51].

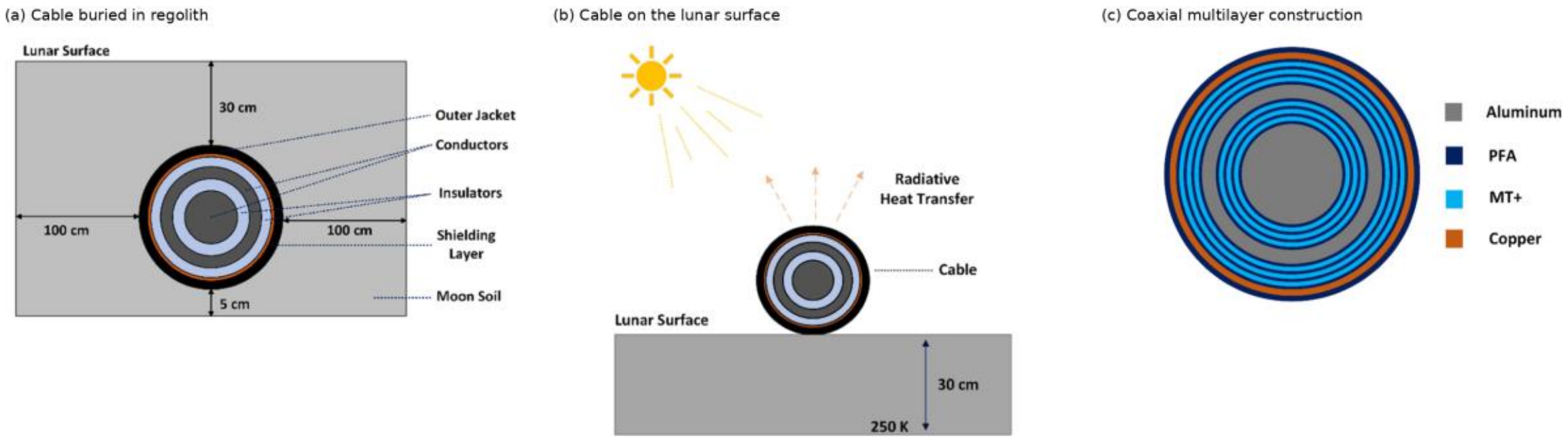

Fig. 27. Lunar coaxial bipolar cable configurations: (a) buried installation, (b) surface installation with radiative exposure, and (c) multilayer insulation and conductor arrangement. Panels reproduced from [51]. The installation boundaries are model assumptions; the drawings are not to a common scale.

### *H.2. Conductor material and installation tradeoffs*

The journal results show a tradeoff between cable mass and occupied cross section. For the buried 1 MW, ±5 kV case, the aluminum cable had a reported mass of 0.9911 kg/m and cross section of 372.02 mm²; the copper cable weighed 1.8240 kg/m and occupied 247.19 mm². Aluminum was therefore approximately 45.7% lighter, but required approximately 50.5% more cable cross section. Its reported J was 368.6941 g·mm, compared with 450.8746 g·mm for copper. Table V gives two buried-cable comparisons using mass per unit length in kg/m, rather than volumetric density [51].

TABLE V. SELECTED BURIED COAXIAL CABLE COMPARISONS AT ±5 kV [51]

| Power | Conductor | Cable mass (kg/m) | Cable area (mm²) | Reported *J* (g·mm) |
|---|---|---|---|---|
| 500 kW | Aluminum | 0.3520 | 133.09 | 46.8410 |
| 500 kW | Copper | 0.5879 | 91.048 | 53.5289 |
| 1 MW | Aluminum | 0.9911 | 372.02 | 368.6941 |
| 1 MW | Copper | 1.8240 | 247.19 | 450.8746 |

The preferred material was not identical for every power, voltage, and installation. At the lower-power buried 100 kW, ±5 kV point, the reported J for aluminum was 1.3221 g·mm, slightly above copper's 1.2117 g·mm, even though aluminum was lighter. At higher power, its mass advantage more often outweighed the area penalty. Increasing the bus voltage reduced current for the same power, but also required more inter-pole insulation. The lowest-mass construction therefore did not automatically give the smallest cross section or the lowest combined metric across all cases [51].

Installation also changed the thermal ranking. In the journal's 1 MW, ±5 kV aluminum comparison, the reported mass was 0.1780 kg/m on the surface, compared with 0.9911 kg/m when buried. The surface cable could reject heat radiatively, whereas the buried design was constrained by conduction through low-conductivity regolith. This modeled mass advantage does not establish that exposed routing is preferable overall: burial changes solar exposure and protection from the lunar environment, while the thermal result remains sensitive to contact, radiative properties, and the prescribed temperature boundaries [51].

### *H.3. Thermal history and relevance to aircraft cable design*

The 2 MW, ±10 kV CEIDP study illustrates the importance of thermal history. A single daytime simulation gave a cable diameter of 20.77 mm, mass of 0.9153 kg/m, and $J$ of 310.11 g·mm. Under the extended 2000 h cyclic calculation, these values increased to 21.29 mm, 0.9620 kg/m, and 342.56 g·mm, respectively. Thus, the reported mass increased by approximately 5.1% and J by approximately 10.5%. Under those modeled conditions, a short initial calculation would have understated the required cable size because of slow heat transfer through the regolith [50].

The lunar journal study also adjusted insulation layer counts to control the calculated electric stress, using a layer-field criterion of 20% of the adopted dielectric strength. Its conductivity calculations used surrogate polymer data, including ETFE data to represent PFA. PFA and ETFE are distinct materials, so those substitutions limit the quantitative interpretation of the electric-field predictions. The results establish numerical design comparisons, rather than experimental qualification for lunar service [51].

For the aircraft review, the transferable outcome is the design method: conductor material, insulation architecture, voltage, heat-rejection path, and thermal history must be assessed together. Lunar vacuum results do not supply an aircraft altitude derating factor, but they show how strongly a preferred cable design can change when the surrounding thermal boundary changes. The combined aircraft and lunar studies provide candidate designs for fabrication and experimental testing.

## V. Fabrication and Experimental Evaluation of MVDC Cable Insulation

The numerical studies in Sections III and IV identified lightweight cable constructions, but their predicted thermal and electrical performance depended on an insulation system that could be manufactured with sufficiently consistent interfaces. Our experimental work therefore began with flat multilayer specimens, progressed to short wrapped cable prototypes, and then examined partial discharge (PD) and breakdown under atmospheric and reduced air pressure. The conference papers document the successive fabrication and measurement stages; the journal papers provide the principal experimental evidence discussed below. Throughout this progression, a fabricated insulation specimen, a complete cable construction, and an aircraft-qualified cable must be distinguished.

### *A. Flat specimens and optimization of the bonding process*

The first fabrication studies established a practical route for assembling thin polymer films into multilayer multifunctional electrical insulation (MMEI). The initial TPEC and DCAS conference papers introduced flat-sample preparation and its optimization [53], [54]. The subsequent IEEE Access study examined the fabrication process systematically using PD measurements and high-voltage testing [55]. Flat specimens provided accessible, repeatable layups before introducing the additional complications of conductor curvature, strand grooves, and wrapping tension.

The investigated constructions combined Kapton MT+ polyimide films with perfluoroalkoxy (PFA) films, which acted as the thermally activated bonding layers. The materials and arrangements corresponded to the selected MMEI architecture, including additional functional layers where required. For the flat electrical specimens, circular copper electrodes were embedded within the polymer layup to avoid the concentrated corner fields associated with square electrodes. Fig. 28 illustrates the layer arrangement and the tooling used to compress and bond it [55].

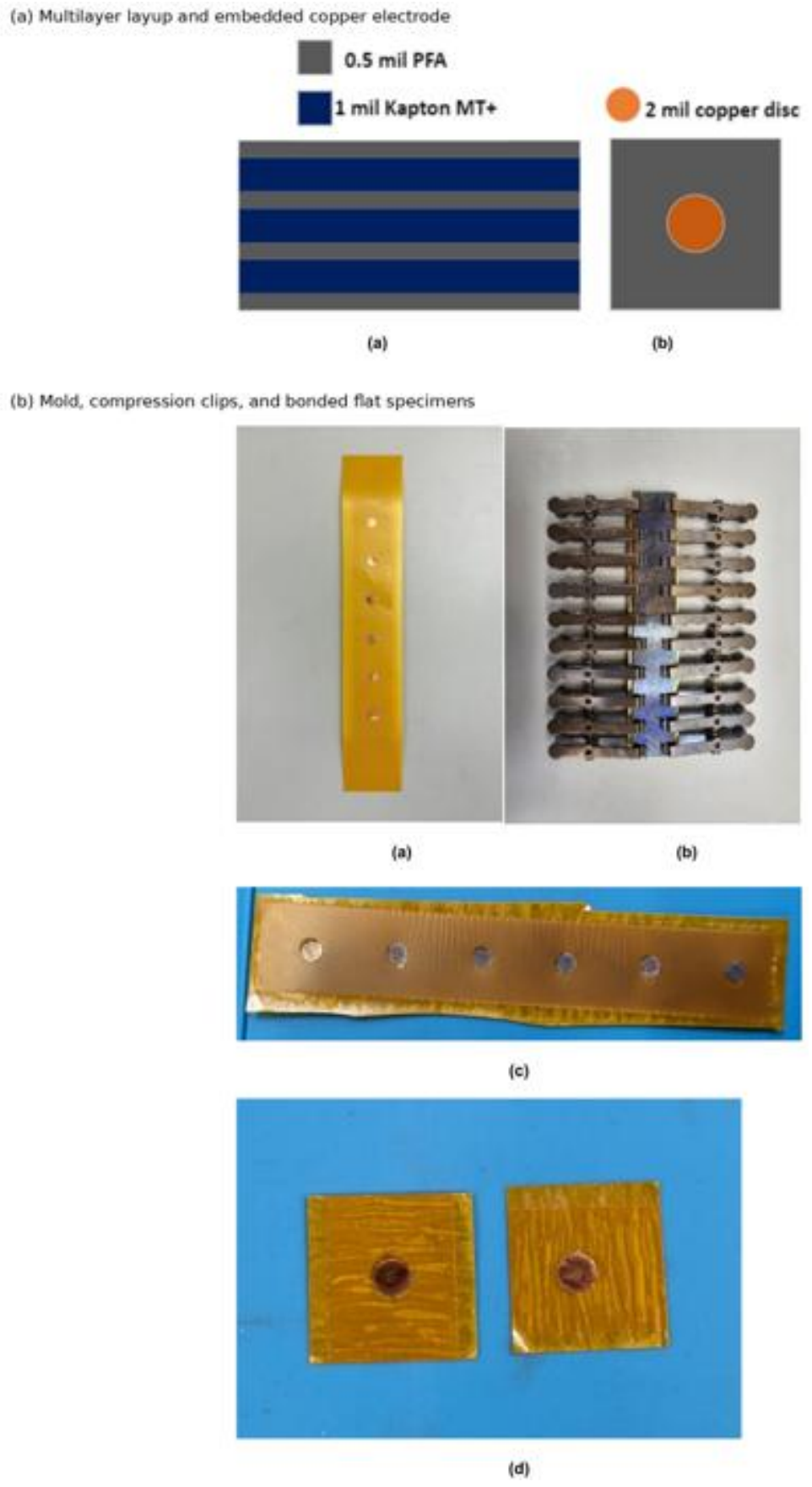

Fig. 28. Fabrication of flat MMEI specimens: (a) layer arrangement and embedded circular electrode; (b) mold, compression clips, and bonded specimens. Panels reproduced from [55].

The fabrication sequence comprised cleaning the molds and films, drying, arranging the films in a steel mold, applying compression with high-temperature clips, heating to fuse the PFA bonding layers, cooling, and cutting specimens. Twelve processing combinations were evaluated: bonding at 320, 340, or 360 °C for either 20 or 40 min, with or without a preceding 250 °C, 2 h vacuum-drying step. The best-performing process in the reported comparison used vacuum predrying followed by bonding at 360 °C for 40 min [55]. This result established a processing recipe for the investigated films and tooling, rather than a universal bonding condition for every insulation construction.

PD measurements also revealed the importance of the fixture. Flat specimens were tested in silicone oil to suppress external discharges, but the oil/electrode arrangement itself exhibited measurable PD, with inception depending on electrode spacing. Consequently, the study considered discharge extinction and pulse magnitude together with inception voltage when comparing processing conditions. The optimized specimens reached 30 kV at atmospheric pressure and 22 kV at reduced pressure without specimen breakdown; the reported limits arose from the test equipment and external air breakdown, respectively. These voltages are demonstrated test levels, not measured intrinsic breakdown strengths [55].

### *B. Transition from flat insulation to wrapped cable specimens*

The KPEC fabrication paper [56] and its expanded Energies study [57] transferred the multilayer process to a cylindrical substrate. Early specimens used a 17.75 mm-diameter, approximately 10 cm-long hollow aluminum tube, making it possible to investigate wrapping and bonding without the surface irregularities of a stranded conductor. The inner insulation employed available Kapton MT film in place of the specified Kapton MT+, together with PFA bonding films. A shortened copper outer electrode was added for electrical testing. These were manufacturing and insulation-screening specimens; they did not reproduce every layer or material of the complete optimized cable [57]. The wrapped layers were compressed with T-bolt hose clamps, vacuum predried at 250 °C for 2 h, and bonded at 360 °C for 40 min. Rounded electrode ends helped reduce external field concentration. Tests on five specimens, with repeated measurements, showed lower PD inception and repetitive inception voltages and greater discharge activity at 18.8 kPa than at 101.325 kPa. The specimen-to-specimen differences also exposed the practical sensitivity of a hand-wrapped insulation system to local bonding quality and fabrication consistency [57].

The subsequent work moved to a commercially available stranded ACSR Linnet conductor with an approximately 170 mm² cross section. The CEIDP pressure study, IPMHVC prototyping study, and IPMHVC dielectric-strength study documented intermediate stages [58]–[60]; the expanded TDEI paper brought fabrication, PD behavior, pressure effects, and breakdown testing together [61]. The stranded conductor was an experimental platform, rather than evidence that its steel-reinforced construction was the final minimum-mass aircraft conductor.

Two wrapping approaches were compared: placing a stacked film arrangement around the conductor, and spirally applying approximately 12.7 mm-wide films with successive layers wound in opposite directions. Tight winding and compression were intended to improve interlayer contact and reduce unbonded regions over the strand surface. The selected thermal process was retained, and protective films separated the insulation from the compression clamps. Fig. 29 shows the compressed layup and finished short cable specimen [61].

Under the reported DC test cycle, spiral wrapping increased PDIV from 2.62 to 3.02 kV and repetitive PDIV (RPDIV) from 3.02 to 3.84 kV. The mean total pulse count over the 23 min cycle decreased from 407.6 to 121.8, while mean pulse charge decreased from 132.96 to 75.03 pC. Fig. 30 compares the measured PDIV, RPDIV, and total pulse counts for the two wrapping methods, illustrating the improvement obtained with spiral wrapping. These measurements support improved fabrication quality for the spiral method under the tested conditions. They do not establish a PD-free 5 kV operating rating, because discharge inception still occurred below that voltage [61].

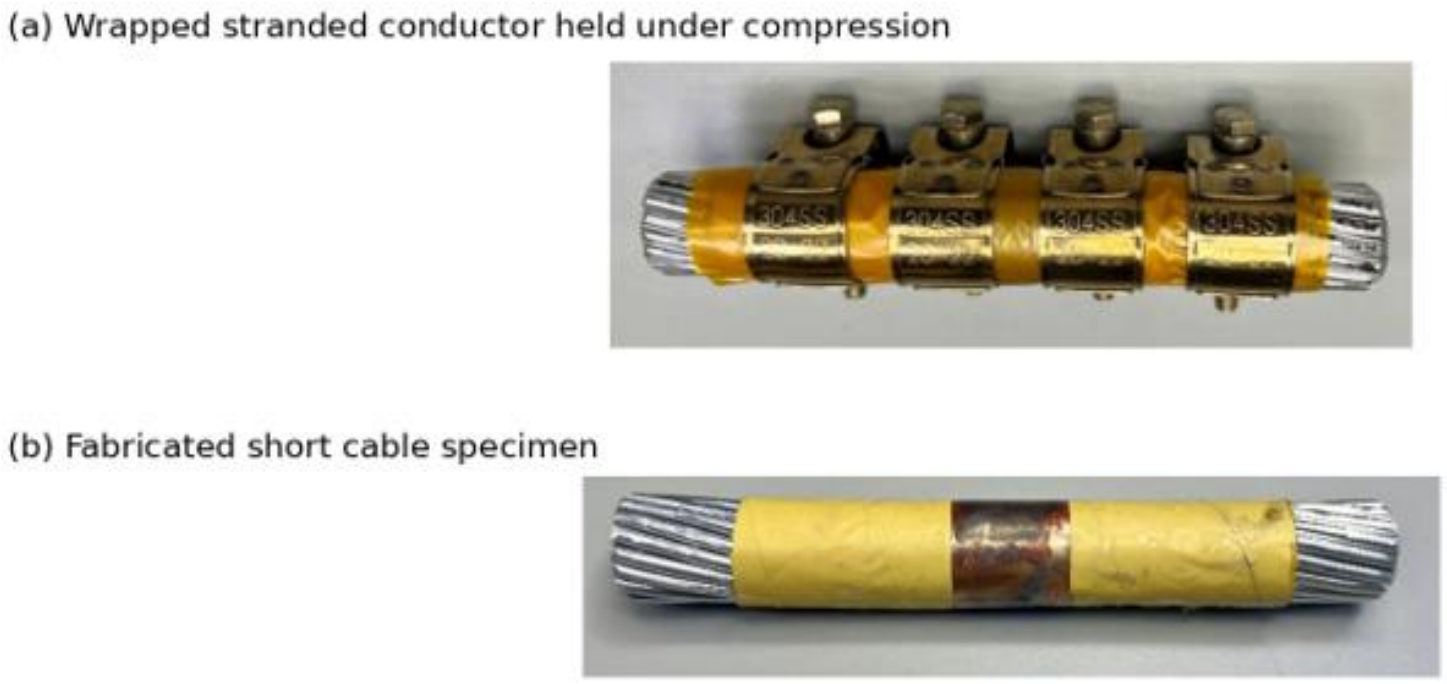

Fig. 29. Transition to a stranded-conductor cable specimen: (a) wrapped ACSR conductor under T-bolt clamp compression and (b) fabricated cable specimen with the test electrode [61].

*C. Pressure-dependent PD testing and interpretation*

The experimental arrangement combined a high-voltage source, coupling capacitor, OMICRON MPD 800 measurement system, and a vacuum chamber containing the specimen (Fig. 31). Reduced-pressure tests represented exposure of the cable installation to unpressurized aircraft compartments. The early cable study used a negative DC ramp from −1 to −5 kV at approximately 1 kV/min, followed by a 5 min hold and a return ramp; later experiments extended the maximum magnitude to 10 kV. Comparisons must therefore retain the applied waveform, voltage limit, sample geometry, and pressure of each study [57], [61]. A 15 pC threshold was used to identify measurable PD in the reported measurements, and RPDIV was defined using a pulse repetition criterion of 20 pulses/min. PDIV identifies the onset of detectable activity, RPDIV identifies sustained repetitive activity under the selected criterion, and PDEV identifies extinction during voltage reduction. These quantities depend on the measurement procedure; they cannot be treated as fixed material constants independent of ramp rate, detection threshold, dwell time, or test fixture [57], [61].

The TDEI pressure sweep on flat specimens showed a general decrease in PDIV and an increase in pulse count and magnitude as air pressure decreased. RPDIV and PDEV exhibited weaker pressure trends. Pulse-sequence analysis further indicated that much of the recorded activity occurred during voltage ramps rather than the steady-voltage hold. This distinction matters for MVDC service: charging transients and voltage changes may initiate activity that is less evident during a short constant-voltage interval. Reduced pressure also changes the discharge conditions in exposed air regions and interfaces, but the chamber pressure alone does not establish the pressure inside a sealed internal void [61].

*D. AC/DC breakdown strength and the influence of specimen geometry*

The GESS conference paper compared the electrical behavior of flat specimens under AC and DC at atmospheric pressure [62]. The expanded TPS study added reduced-pressure measurements and wrapped cable specimens, permitting pressure, waveform, and specimen geometry to be examined together [63]. Unlike the nonfailure test limits in Section V.A, these experiments included destructive breakdown measurements and reported two-parameter Weibull fits. The characteristic strength is the fitted 63.2% breakdown quantile for the tested population; it is not an allowable continuous operating field.

Values are Weibull characteristic strengths, not cable voltage ratings. Percentage reductions are calculated relative to atmospheric-pressure values. AC and DC results retain the voltage conventions of [63].

Table VI shows that pressure sensitivity differed substantially with waveform and geometry. The wrapped specimens had lower characteristic strength than flat specimens under the corresponding conditions, particularly under AC at 18.8 kPa. The result reinforces the need to test the fabricated cable geometry: a flat-film result does not capture strand interfaces, wrapping defects, curvature, or electrode-end conditions. It also shows why one pressure derating factor should not be transferred indiscriminately between AC and DC or between flat and wrapped insulation [63].

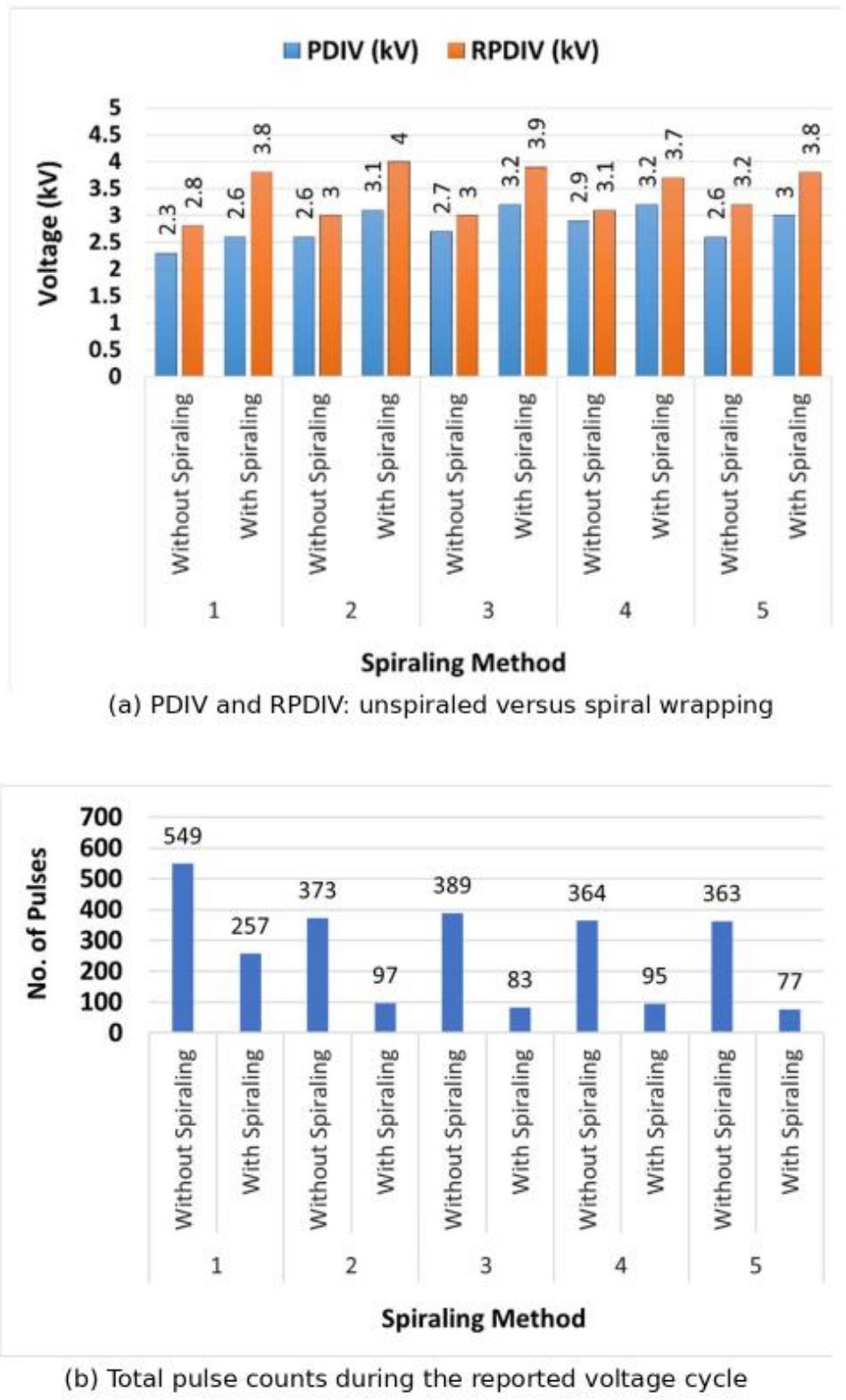

Fig. 30. Selected evidence for the wrapping-method improvement: (a) PDIV and RPDIV and (b) pulse count during the reported voltage cycle [61]. Repeated measurements characterize the tested specimens; they do not represent additional independently manufactured cables.

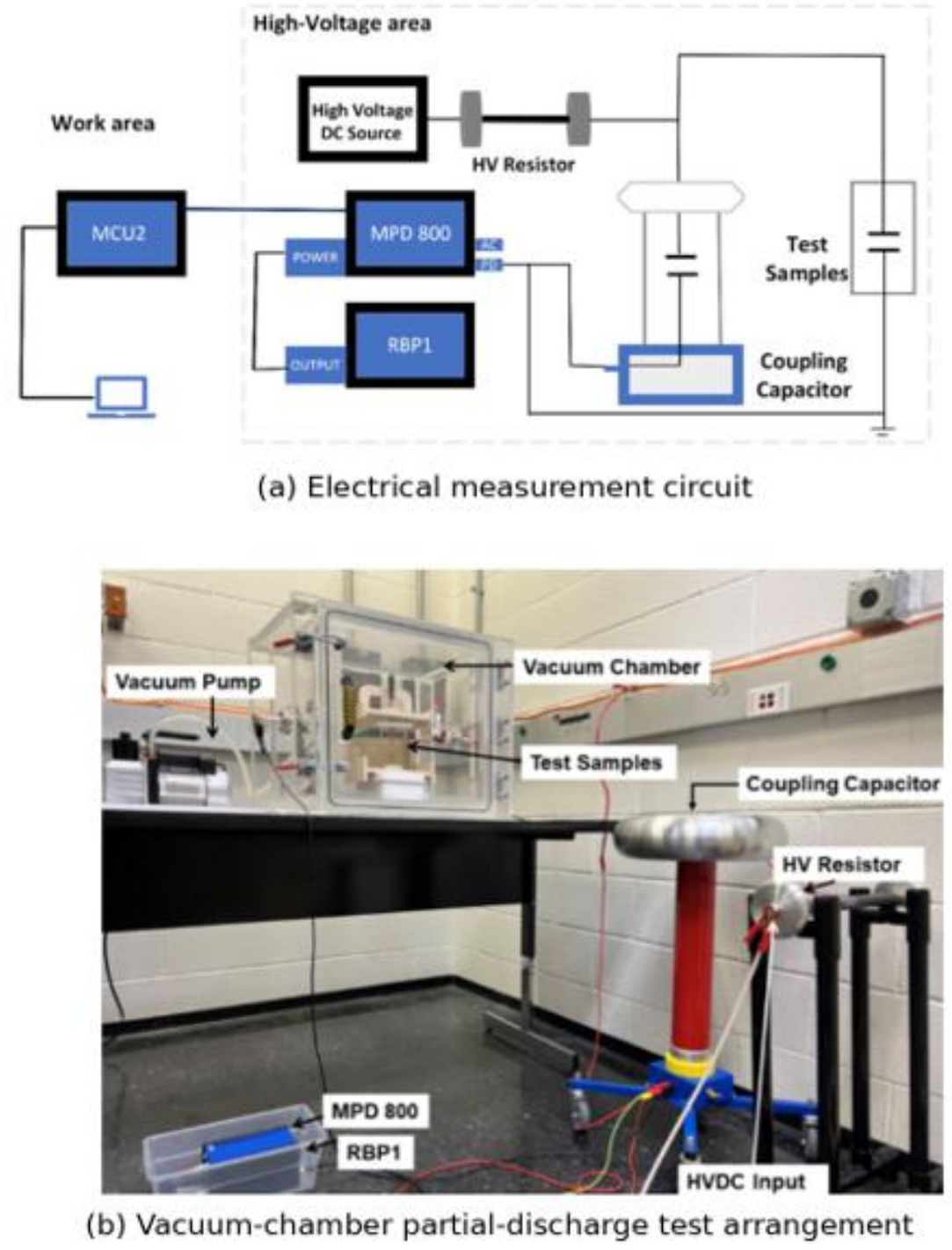

Fig. 31. PD measurement arrangement: (a) electrical circuit and (b) vacuum-chamber experimental setup. Panels reproduced from [61].

TABLE VI. Reported characteristic breakdown strengths of MMEI specimens [63]

| Specimen | Voltage | 101.325 kPa (kV/mm) | 18.8 kPa (kV/mm) | Reduction |
|---|---|---|---|---|
| Flat | DC | 143.52 | 107.33 | 25.2% |
| Flat | AC | 96.81 | 93.48 | 3.4% |
| Wrapped cable | DC | 112.60 | 78.67 | 30.1% |
| Wrapped cable | AC | 87.10 | 38.72 | 55.5% |

The preceding TDEI study reported flat-specimen DC characteristic strengths of 143.52 kV/mm at 101 kPa and 108.74 kV/mm at 21 kPa [61]. These are consistent in trend with the later comparison, but its 21 kPa result must not be relabeled as an 18.8 kPa measurement. Breakdown fields normalized by measured insulation thickness and breakdown voltages across a complete specimen are also distinct quantities; both geometry and the thickness definition must accompany comparisons [61], [63].

### *E. Comparison with a conventional medium-voltage cable*

The next stage asked whether the MMEI prototypes offered an experimental advantage over an available conventional cable. The 2025 KPEC paper addressed PD [64], and the SDEMPED paper addressed dielectric strength [65]. Their expanded TPS journal study compared both properties at atmospheric pressure and 18.8 kPa [66]. The comparison used a 5 kV-rated conventional underground cable with a 500 kcmil stranded copper conductor, approximately 2.921 mm of EPR insulation, and a PVC jacket, alongside the MMEI specimen built on the smaller ACSR conductor.

At atmospheric pressure, the reported PDIV values were approximately 2.3–2.7 kV for MMEI and 1.7–1.8 kV for the conventional cable. At 18.8 kPa, the respective ranges were approximately 1.4–1.6 and 1.3–1.8 kV, so the ranges overlapped. The study observed extinction in the MMEI specimens under its procedure but did not record PDEV for the conventional specimens at reduced pressure. This is a measurement outcome under that voltage cycle, rather than proof that discharge can never extinguish in a conventional cable [66].

The MMEI specimens also exhibited substantially higher thickness-normalized breakdown strength in the reported comparison. However, conductor material and dimensions, screens, insulation thickness, and overall construction differed between the two cable types. Accordingly, these results compare tested cable systems rather than isolating an intrinsic advantage of one polymer over another. External air flashover and termination conditions must also be distinguished from bulk insulation breakdown before generalizing the findings to installed cables. The experiments support further development of the MMEI approach; they do not establish that all conventional cable constructions are unsuitable for aircraft [66].

### *F. Insulation replacement within a common cable platform*

A subsequent manuscript sought to reduce the construction differences by retaining the conventional cable platform and replacing its primary insulation with MMEI [67]. Fig. 32 identifies the conductor, conductor screen, primary insulation, insulation screen, copper tape shield, and PVC jacket in the baseline construction. The reported redesign retained the other cable components while reducing nominal primary-insulation thickness from 115 to 11.5 mil, approximately 2.92 to 0.292 mm. It therefore compares an insulation-and-thickness redesign within a common platform; it is not a constant-thickness material substitution. These results extend the earlier comparison by reducing differences in conductor and shielding construction.

Five conventional specimens and two MMEI specimens were evaluated. Repeated PD measurements on a specimen do not increase the number of independently manufactured specimens. Table VII summarizes the measured breakdown voltages. At 18.8 kPa, the two MMEI specimens broke down at 13.50 and 14.69 kV, compared with 4.42–4.89 kV for the conventional specimens. These tabulated measurements are the basis of the comparison here [67].

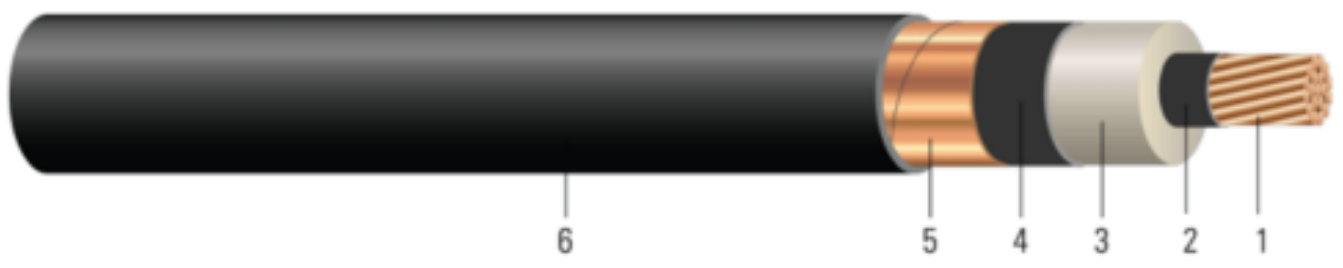

Fig. 32. Conventional cable platform used for the insulation-replacement comparison. The primary-insulation region was replaced by the thinner MMEI construction while the other cable components were retained as described in [67]. Labels: (1) copper conductor, (2) conductor screen, (3) primary insulation, (4) insulation screen, (5) copper tape shield, and (6) PVC jacket. Illustration reproduced from [67].

Ranges summarize the tabulated measurements in [67]. The nominal insulation thickness also changed, so this is a cable redesign comparison. Two MMEI specimens do not establish production variability or a reliable breakdown probability distribution.

TABLE VII. Breakdown voltages in the common-platform insulation comparison [67]

| Cable construction | Independent specimens | 101.325 kPa (kV) | 18.8 kPa (kV) |
|---|---|---|---|
| Conventional insulation | 5 | 5.58–8.26 | 4.42–4.89 |
| MMEI insulation replacement | 2 | 23.30–24.17 | 13.50–14.69 |

The PD results were more nuanced than the breakdown results. Mean atmospheric-pressure PDIV increased from approximately 1.74 to 2.20 kV, whereas mean reduced-pressure PDIV changed only from approximately 1.46 to 1.50 kV. PDEV was reported for the MMEI specimens at 18.8 kPa but was not recorded for the conventional specimens during the same procedure. The strong improvement in measured breakdown voltage therefore should not be described as a comparable improvement in low-pressure PD inception. Both quantities matter: a cable may survive a short high-voltage test while exhibiting PD at much lower voltage [67].

### *G. Outcome of the fabrication and test program*

Taken together, these studies established a laboratory fabrication route from bonded flat films to short stranded-conductor cable prototypes, identified a processing window and wrapping method that reduced measured discharge activity, and demonstrated that pressure, waveform, geometry, and complete cable construction materially affect the electrical results. The journal papers provide the main evidence, while the conference publications document the successive steps leading to those investigations [53]–[67]. The experiments also identify the limits of the present validation: not every prototype reproduced the complete simulated construction, and measurable PD often began below the intended 5 kV pole-to-ground operating level.

The ±5 kV, 1 kA objective established in Section II thus remains a combined design and qualification target. The electrical measurements presented here do not, by themselves, demonstrate 1 kA ampacity, long-duration PD-free operation, or flight qualification. The next development step is to combine the selected complete insulation construction with representative terminations and installation boundaries, and verify its electrical and thermal behavior under the relevant current, pressure, temperature, and mission-duration conditions. That distinction connects the numerical design results to the demonstrated prototypes without treating short-specimen electrical tests as validation of every aircraft service requirement.

## VI. Conclusions

This article has traced our research journey toward MVDC power cables for future wide-body all-electric aircraft, connecting aircraft-level requirements with modeling, design, fabrication, and experimental evaluation. Its focus on publications from this research program documents how successive investigations addressed the linked challenges of propulsion-power delivery, cable mass, heat rejection, and insulation performance under reduced pressure. The system-level studies established ±5 kV distribution and a 1 kA design objective for the most heavily loaded circuits under the adopted contingency conditions. Subsequent investigations introduced flight-phase requirements, distinguishing the higher-current takeoff condition from lower-current cruise operation at 18.8 kPa. These studies showed that cable sizing must account for current, pressure, temperature, installation conditions, and mission duration together. The coupled models identified promising multilayer insulation systems and demonstrated the influence of conductor geometry, pole separation, duct configuration, and cooling method on cable mass and dimensions. They also showed that minimizing cable mass alone does not necessarily minimize the mass or occupied space of the complete installation. Fabrication studies translated selected concepts into bonded flat specimens and short wrapped cable prototypes. Controlled drying and bonding, together with improved wrapping methods, reduced measured discharge activity. Electrical testing demonstrated that pressure, waveform, specimen geometry, and fabrication quality substantially influence partial-discharge and breakdown behavior. The contribution of this research program is the integrated progression from system requirements to noncryogenic MVDC cable designs targeting ±5 kV and 1 kA, followed by prototype fabrication and electrical testing at pressures down to 18.8 kPa. The numerical ratings and experimental achievements represent complementary stages of development; they do not yet establish simultaneous operation at the full target voltage and current under reduced pressure. Further work should verify complete cable constructions and representative terminations through combined current–voltage testing, mission-relevant environmental cycles, long-duration insulation endurance, and mechanical evaluation. These steps will connect the demonstrated laboratory advances to reliable aircraft cable installations.

## References

[1] S. Chowdhury and M. Ghassemi, "Next-gen aviation: Who will rule the skies—Hydrogen or Electric? — A review," *IEEE Access*, vol. 13, pp. 155141-155154, 2025.
[2] Airbus, "A320neo," [Online]. Available: https://www.aircraft.airbus.com/en/aircraft/a320-family/a320neo

[3] Boeing, "737 MAX," [Online]. Available: https://www.boeing.com/commercial/737max
[4] Boeing, "787 Dreamliner," [Online]. Available: https://www.boeing.com/commercial/787
[5] Airbus, "Airbus A350 family: The future of long-haul travel," [Online]. Available: https://www.airbus.com/en/products-services/commercial-aircraft/passenger-aircraft/a350-family
[6] Federal Aviation Administration, "ASE performance above FL410," Altimetry System Error Technical Interchange Meeting, Sept. 2025, slide 3. [Online]. Available: https://www.faa.gov/NAARMO_2025_ASE_TIM_US_ASE_FlightLevels_Above_410
[7] A. Barzkar and M. Ghassemi, "Electric power systems in more and all electric aircraft: a review," *IEEE Access*, vol. 8, pp. 169314–169332, Sept. 2020.
[8] A. Barzkar and M. Ghassemi, "Components of electrical power systems in more and all-electric aircraft: A review," *IEEE Trans. Transport. Electrific.*, vol. 8, no. 4, pp. 4037-4053, Dec. 2022.
[9] Airbus, "Airbus and MTU Aero Engines to create a joint venture to develop a fully electric hydrogen fuel cell engine," July 7, 2026. [Online]. Available: https://www.airbus.com/en/newsroom/press-releases/2026-07-airbus-and-mtu-aero-engines-to-create-a-joint-venture-to-develop-a-fully-electric-hydrogen-fuel-cell?trk=article-ssr-frontend-pulse_little-text-block&utm_source=chatgpt.com
[10] NASA, "NASA, GE Aerospace Work Enables Hybrid-Electric Flight Demonstration," July 20, 2026. [Online]. Available: https://www.nasa.gov/directorates/rtmd/nasa-ge-hybrid-electric-flight/?utm_source=chatgpt.com
[11] T. Clynes, "How Beta Technologies is clearing the runway for electric flight," *IEEE Spectrum*, pp. 46-53, Nov. 2026.
[12] S. Karpuk, Y. Freund, and R. Hanke-Rauschenbach, "Potential of hydrogen fuel cell aircraft for commercial applications with advanced airframe and propulsion technologies," *Aerospace*, vol. 12, no. 1, art. 35, 2025.
[13] NOAA, NASA, and U.S. Air Force, U.S. Standard Atmosphere, 1976, NOAA-S/T 76-1562 and NASA-TM-X-74335, Oct. 1976. [Online]. Available: https://ntrs.nasa.gov/citations/19770009539
[14] P. Vrchota, A. Prachař, M. Šmíd, and J. Middel, "Contribution of conformal antennas towards sustainable aircraft," in Proc. Eur. Conf. Multifunctional Structures (EMuS), Barcelona, Spain, Jun. 2019, pp. 38–47, Fig. 7. [Online]. Available: https://congress.cimne.com/emus2019/frontal/doc/EbookEMuS2019.pdf
[15] M. Ghassemi and A. Barzkar, "Power flow solvers for medium voltage direct current (MVDC) microgrids," in *Proc. 6th IEEE Workshop Electron. Grid (eGRID)*, 2021, pp. 1–6, doi: 10.1109/EGRID52793.2021.9662135.
[16] M. Ghassemi and A. Barzkar, "DC load flow models for the electric power system of wide body all electric aircraft," in *Proc. IEEE Aerospace Conf. (AERO)*, 2022, pp. 1–8, doi: 10.1109/AERO53065.2022.9843788.
[17] M. Ghassemi and M. Saghafi, "Optimal electric power system architectures for wide body all electric aircraft," in *Proc. IEEE Aerospace Conf. (AERO)*, 2022, pp. 1–9, doi: 10.1109/AERO53065.2022.9843195.
[18] M. Ghassemi, A. Barzkar, and M. Saghafi, "All-electric NASA N3-X aircraft electric power systems," *IEEE Trans. Transport. Electrific.*, vol. 8, no. 4, pp. 4091–4104, Dec. 2022, doi: 10.1109/TTE.2022.3158186.
[19] A. Saha and M. Ghassemi, "Electrical system architecture," in Aviation Electrification, T. Yang, B. Sarlioglu, and P. Wheeler, Eds. Elsevier, forthcoming, 2027.
[20] J. L. Felder, M. T. Tong, and J. Chu, "Sensitivity of mission energy consumption to turboelectric distributed propulsion design assumptions on the N3-X hybrid wing body aircraft," in *Proc. AIAA Joint Propulsion Conf.*, Atlanta, GA, USA, Jul. 2012, NASA NTRS 20130010734. [Online]. Available: https://ntrs.nasa.gov/citations/20130010734
[21] NASA, "N3-X," May 5, 2026. [Online]. Available: https://www.nasa.gov/eap-aircraft-concepts/n3-x/
[22] NASA, "Advanced Aircraft Concepts for Environmental Sustainability (AACES) 2050." [Online]. Available: https://www.nasa.gov/reference/advanced-aircraft-concepts-for-environmental-sustainability-2050/
[23] Boeing, "787 Dreamliner by design," [Online]. Available: https://www.boeing.com/commercial/787/by-design
[24] M. Ghassemi, "Design of reliable and resilient electric power systems for wide-body all-electric aircraft," in Design Reliability and Maintainability of Energy Systems, F. P. García Márquez, R. V. Sánchez Loja, and M. Papaelias, Eds., 1st ed. Elsevier, 2026, ch. 9. ISBN: 9780443452314; eBook ISBN: 9780443452321.
[25] M. Borghei and M. Ghassemi, "Insulation materials and systems for more- and all-electric aircraft: A review identifying challenges and future research needs," IEEE Trans. Transport. Electrific., vol. 7, no. 3, pp. 1930–1953, Sept. 2021, doi: 10.1109/TTE.2021.3050269.

[26] A. Saha and M. Ghassemi, "A review of electrical insulation challenges for electrical powertrain components in the future more- and all-electric aircraft under low-pressure conditions," IEEE J. Emerg. Sel. Topics Power Electron., vol. 13, no. 4, pp. 4521–4536, Aug. 2025, doi: 10.1109/JESTPE.2024.3491298.
[27] A. Saha and M. Ghassemi, "Challenges in electrical insulation materials and thermal management for medium voltage power cables for envisaged wide-body all-electric aircraft," in Aeronautics—Characteristics and Emerging Technologies, L. Li, Ed. London, U.K.: IntechOpen, 2024, doi: 10.5772/intechopen.1006159.
[28] Y. Ohki et al., "Advances in polymeric insulating materials: Global progress in these 10 years," High Voltage, vol. 11, no. 5, pp. 1213–1249, 2026, doi: 10.1049/hve2.70257.
[29] S. Chowdhury and M. Ghassemi, "Medium-voltage power cables for electrified aircraft and spacecraft: Insulation physics, environmental stresses and qualification gaps," High Voltage, 2026, doi: 10.1049/hve2.70267.
[30] A. Azizi, M. Ghassemi, and J. M. Lehr, "Influence of low pressure on thermal limit of MVDC power cables used in all electric aircraft," in *Proc. IEEE Int. Power Modulator High Voltage Conf. (IPMHVC)*, Knoxville, TN, USA, 2022, pp. 88–91, doi: 10.1109/IPMHVC51093.2022.10099406.
[31] A. Azizi, M. Ghassemi, and J. M. Lehr, "Heat transfer challenges for MVDC power cables used in wide body all electric aircraft under low pressures," *IEEE Access*, vol. 10, pp. 111811–111819, 2022, doi: 10.1109/ACCESS.2022.3216300.
[32] A. Azizi, M. Ghassemi, and J. M. Lehr, "Design of a cable system for a high-power density MVDC aircraft electric power system," in *Proc. IEEE Conf. Electr. Insul. Dielectric Phenomena (CEIDP)*, Denver, CO, USA, 2022, pp. 151–154, doi: 10.1109/CEIDP55452.2022.9985375.
[33] A. Azizi and M. Ghassemi, "Design of high power density MVDC cables for wide-body all electric aircraft," *IEEE Trans. Dielectr. Electr. Insul.*, vol. 30, no. 5, pp. 2315–2324, Oct. 2023, doi: 10.1109/TDEI.2023.3285849.
[34] F. Islam and M. Ghassemi, "NASA N3-X aircraft DC power system design," in *Proc. IEEE 17th Dallas Circuits Syst. Conf. (DCAS)*, 2024, pp. 1–5, doi: 10.1109/DCAS61159.2024.10539897.
[35] F. Islam, A. Saha, M. A. Rahman, and M. Ghassemi, "Bipolar MVDC cable under different ambient conditions for all-electric aircraft," in *Proc. IEEE Kansas Power Energy Conf. (KPEC)*, 2024, pp. 1–4, doi: 10.1109/KPEC61529.2024.10676036.
[36] F. Islam and M. Ghassemi, "Optimal design of MVDC power cables for different air pressures for wide-body all-electric aircraft," *IEEE Access*, vol. 12, pp. 185937–185945, 2024, doi: 10.1109/ACCESS.2024.3514825.
[37] A. Saha, A. Azizi, and M. Ghassemi, "An optimal bipolar MVDC coaxial power cable design for envisaged all electric wide body aircraft," in *Proc. IEEE Conf. Electr. Insul. Dielectric Phenomena (CEIDP)*, East Rutherford, NJ, USA, 2023, pp. 1–4, doi: 10.1109/CEIDP51414.2023.10410570.
[38] A. Azizi, A. Saha, and M. Ghassemi, "A cuboid geometry design for MVDC power cables for using in future all electric wide body aircraft," in *Proc. IEEE Conf. Electr. Insul. Dielectric Phenomena (CEIDP)*, East Rutherford, NJ, USA, 2023, pp. 1–4, doi: 10.1109/CEIDP51414.2023.10410513.
[39] A. Saha, A. Azizi, and M. Ghassemi, "Optimal bipolar MVDC power cable designs for future wide-body all electric aircraft," *IEEE Trans. Dielectr. Electr. Insul.*, vol. 31, no. 4, pp. 2074–2083, Aug. 2024, doi: 10.1109/TDEI.2024.3355033.
[40] A. Saha and M. Ghassemi, "MVDC bipolar power cables with rectangular geometry design for envisaged all-electric wide-body aircraft," in *Proc. IEEE Texas Power Energy Conf. (TPEC)*, College Station, TX, USA, 2024, pp. 1–5, doi: 10.1109/TPEC60005.2024.10472255.
[41] A. Saha and M. Ghassemi, "Optimized design of ±5 kV, 1 kA rectangular power cable for a low pressure of 18.8 kPa for envisioned all-electric wide-body aircraft," *IEEE Access*, vol. 12, pp. 28654–28665, 2024, doi: 10.1109/ACCESS.2024.3368437.
[42] A. Saha and M. Ghassemi, "Design and environmental performance of rectangular bipolar cables for all-electric wide-body aircraft," *IEEE Trans. Ind. Appl.*, vol. 62, no. 3, pp. 5386–5395, May/Jun. 2026, doi: 10.1109/TIA.2025.3619924.
[43] A. Azizi and M. Ghassemi, "An analysis of parameters affecting ampacity in aircraft bipolar MVDC power cables via coupled electrical, thermal, and computational fluid dynamic modelling," *High Voltage*, vol. 9, no. 6, pp. 1208–1220, 2024, doi: 10.1049/hve2.12452.
[44] A. Saha and M. Ghassemi, "Thermal and electrical performance of multilayer, multifunctional electrical insulated cylindrical and cuboid cables in wide-body all-electric aircraft," in *Proc. IEEE Symp. Diagnostics Electr. Mach., Power Electron. Drives (SDEMPED)*, 2025, pp. 1–6, doi: 10.1109/SDEMPED53223.2025.11153972.
[45] A. Saha and M. Ghassemi, "Optimal duct design for MVDC cables in all-electric wide-body aircraft," in *Proc. IEEE Kansas Power Energy Conf. (KPEC)*, 2025, pp. 1–5, doi: 10.1109/KPEC65465.2025.11045041.

[46] A. Saha and M. Ghassemi, “Design and analysis of medium voltage DC cable ducts for electric aircraft considering thermal and electrical insulation requirements,” *IEEE Trans. Plasma Sci.*, vol. 53, no. 8, pp. 2058–2067, Aug. 2025, doi: 10.1109/TPS.2025.3586200.
[47] M. A. Rahman, A. Saha, and M. Ghassemi, “Effect of forced heat convection on heat transfer for bipolar MVDC power cables in envisaged wide-body all-electric aircraft,” in *Proc. IEEE 17th Dallas Circuits Syst. Conf. (DCAS)*, 2024, pp. 1–5, doi: 10.1109/DCAS61159.2024.10539872.
[48] M. A. Rahman and M. Ghassemi, “Analysis of forced convection heat transfer for bipolar MVDC power cable envisaged for wide-body all-electric aircraft,” *IEEE Trans. Dielectr. Electr. Insul.*, vol. 33, no. 5, pp. 3801–3810, Oct. 2026, doi: 10.1109/TDEI.2026.3676013.
[49] A. Saha and M. Ghassemi, “High power density, cost-effective HVDC cables for power transmission on the Moon,” in *Proc. IEEE Texas Power Energy Conf. (TPEC)*, 2024, pp. 1–5, doi: 10.1109/TPEC60005.2024.10472192.
[50] A. Saha and M. Ghassemi, “Developing an MVDC bipolar coaxial cable system for lunar power transmission,” in *Proc. IEEE Conf. Electr. Insul. Dielectric Phenomena (CEIDP)*, 2024, pp. 1–4, doi: 10.1109/CEIDP61745.2024.10907587.
[51] A. Saha and M. Ghassemi, “Designing high-power density MVDC bipolar power cables for power transmission on the Moon,” *IEEE Trans. Aerosp. Electron. Syst.*, vol. 61, no. 2, pp. 5079–5087, Apr. 2025, doi: 10.1109/TAES.2024.3513959.
[52] A. Saha and M. Ghassemi, “Optimal design of high-power density medium-voltage direct current bipolar power cables for lunar power transmission,” *Aerospace*, vol. 11, no. 8, art. 685, 2024, doi: 10.3390/aerospace11080685.
[53] A. Azizi, A. Saha, M. A. Rahman, M. Ghassemi, and J. Lehr, “Fabrication of MVDC multilayer insulation systems as flat samples for wide-body all-electric aircraft,” in *Proc. IEEE Texas Power Energy Conf. (TPEC)*, 2024, pp. 1–5, doi: 10.1109/TPEC60005.2024.10472272.
[54] S. Chowdhury, A. Azizi, A. Saha, M. A. Rahman, M. Ghassemi, and J. Lehr, “An optimal approach to fabricate MVDC multilayer insulation systems as flat samples for wide-body all-electric aircraft,” in *Proc. IEEE Dallas Circuits Syst. Conf. (DCAS)*, 2024, pp. 1–5, doi: 10.1109/DCAS61159.2024.10539902.
[55] S. Chowdhury, A. Saha, M. A. Rahman, S. P. Kalakonda, and M. Ghassemi, “Study of partial discharge characteristics of optimally fabricated MMEI flat samples under DC and their DC dielectric strength for applications in envisaged all-electric wide-body aircraft,” *IEEE Access*, vol. 12, pp. 88547–88557, 2024, doi: 10.1109/ACCESS.2024.3418789.
[56] M. A. Rahman, A. Saha, M. Ghassemi, and J. Lehr, “Fabrication optimization and PD characteristics of MVDC multilayer insulation cable systems for envisaged wide-body all-electric aircraft,” in *Proc. IEEE Kansas Power Energy Conf. (KPEC)*, 2024, pp. 1–5, doi: 10.1109/KPEC61529.2024.10676167.
[57] M. A. Rahman, A. Saha, and M. Ghassemi, “Optimized fabrication process and PD characteristics of MVDC multilayer insulation cable systems for next generation wide-body all-electric aircraft,” *Energies*, vol. 17, no. 12, art. 3040, 2024, doi: 10.3390/en17123040.
[58] M. A. Rahman, A. Saha, S. Chowdhury, M. Ghassemi, and J. Lehr, “Analysis of partial discharge characteristics of MVDC power cables optimally designed for all-electric wide body aircraft under varying pressure during takeoff and descent,” in *Proc. IEEE Conf. Electr. Insul. Dielectric Phenomena (CEIDP)*, 2024, pp. 1–4, doi: 10.1109/CEIDP61745.2024.10907339.
[59] A. Saha, M. A. Rahman, S. Chowdhury, M. Ghassemi, and J. Lehr, “Prototyping aircraft MVDC power cables with optimal multilayer multifunctional electrical insulation systems,” in *Proc. IEEE Int. Power Modulator High Voltage Conf. (IPMHVC)*, 2024, pp. 1–4, doi: 10.1109/IPMHVC55105.2024.11002820.
[60] S. Chowdhury, A. Saha, M. A. Rahman, M. Ghassemi, and J. Lehr, “The influence of low pressure on dielectric strength of aircraft MVDC power cable,” in *Proc. IEEE Int. Power Modulator High Voltage Conf. (IPMHVC)*, 2024, pp. 1–4, doi: 10.1109/IPMHVC55105.2024.11002819.
[61] A. Saha, S. Chowdhury, M. A. Rahman, and M. Ghassemi, “Analysis of partial discharge characteristics and dielectric strength in multilayer insulation systems for MVDC cables in future all-electric wide-body aircraft,” *IEEE Trans. Dielectr. Electr. Insul.*, vol. 32, no. 4, pp. 2284–2293, Aug. 2025, doi: 10.1109/TDEI.2025.3557779.
[62] S. Chowdhury, A. Saha, M. A. Rahman, and M. Ghassemi, “A comparative study on partial discharge and dielectric strength of MMEI flat samples under AC and DC conditions at atmospheric pressure,” in *Proc. IEEE Green Energy Smart Syst. Conf. (GESS)*, 2024, pp. 1–5, doi: 10.1109/GESS63533.2024.10785071.
[63] S. Chowdhury, M. A. Rahman, and M. Ghassemi, “Comparative analysis of partial discharge and dielectric strength of multilayer multifunctional electrical insulation (MMEI) samples under AC and DC conditions at atmospheric and low pressure,” *IEEE Trans. Plasma Sci.*, vol. 53, no. 9, pp. 2401–2409, Sept. 2025, doi: 10.1109/TPS.2025.3596603.

[64] M. A. Rahman, S. Chowdhury, and M. Ghassemi, "Partial discharge characteristics comparison between MV underground cable and MMEI cable for envisaged wide-body all-electric aircraft," in *Proc. IEEE Kansas Power Energy Conf. (KPEC)*, 2025, pp. 1–5, doi: 10.1109/KPEC65465.2025.11044913.

[65] M. A. Rahman, S. Chowdhury, and M. Ghassemi, "A comparative study of dielectric strength between MV underground cable and MMEI cable for envisaged wide-body all-electric aircraft," in *Proc. IEEE Symp. Diagnostics Electr. Mach., Power Electron. Drives (SDEMPED)*, 2025, pp. 1–6, doi: 10.1109/SDEMPED53223.2025.11154189.

[66] M. A. Rahman, S. Chowdhury, and M. Ghassemi, "Comparative study of partial discharge characteristics and dielectric strength in multilayer multifunctional electrical insulation (MMEI)-based and conventional MVdc power cables for all electric aircraft," *IEEE Trans. Plasma Sci.*, vol. 54, no. 3, pp. 1180–1191, Mar. 2026, doi: 10.1109/TPS.2026.3656301.

[67] S. Chowdhury and M. Ghassemi, "Performance enhancement of MVDC aircraft cables using micro-multilayer insulation under low-pressure conditions," *arXiv preprint arXiv:2604.10238*, 2026, doi: 10.48550/arXiv.2604.10238.